\documentclass[12pt]{article}
	\usepackage{graphicx}
	\usepackage{graphics}
	\usepackage[margin=.8in]{geometry}
	\usepackage{amsmath,amssymb}
	\usepackage{hyperref,lineno}
	\usepackage{epsfig,epstopdf}
	\usepackage{siunitx}
	\usepackage{upgreek}	
	\usepackage{setspace}
	\usepackage{authblk}
	\usepackage[font=small, labelfont = bf]{subcaption}
	\usepackage[font=small,labelfont=bf]{caption}
	\begin{document}
		
			\title{Investigating the normal and tangential peeling behaviour of gecko spatulae using a coupled adhesion-friction model}
			\author[1]{\small{Saipraneeth Gouravaraju}}
			\affil[1]{\footnotesize{Indian Institute of Technology Guwahati, Guwahati, India 781039}}
			\author[2]{\small{Roger A. Sauer}}
			\affil[2]{\footnotesize{Aachen Institute for Advanced Study in Computational Engineering Science (AICES), RWTH Aachen University, Templergraben 55, 52056 Aachen, Germany}}
			\author[1]{\small{Sachin Singh Gautam}\footnote{Corresponding Author, email: \href{mailto:ssg@iitg.ac.in}{ssg@iitg.ac.in}}}
			\date{}
			\maketitle
			\begin{abstract}
				The present work investigates the normal and tangential peeling behaviour of a gecko spatula using a coupled adhesion-friction model. The objective is to explain the strong attachment and easy detachment behaviour of the spatulae as well as to understand the principles behind their optimum design. Using nonlinear finite element computations, it is shown that during \emph{tangentially-constrained} peeling the partial sliding of the spatula pad near the peeling front stretches the spatula, thus increasing the strain energy and leading to high pull-off forces. The model is used to investigate the influence of various parameters on the pull-off forces -- such as the peeling angle, spatula shaft angle, strip thickness, and material stiffness. The model shows that increasing the spatula pad thickness beyond a certain level does not lead to a significant increase in the attachment forces. Further, the easy detachment behaviour of geckos is studied under \emph{tangentially-free} peeling conditions. It is found that the spatulae readily detach from the substrate by changing their shaft angle and eventually peel vertically like a tape. Since the present computational model is not limited by the geometrical, kinematical, and material restrictions of theoretical models, it can be employed to analyse similar biological adhesive systems. 
			\end{abstract}
			
			
			\section{Introduction}
			The underside of each digit on the gecko toes is covered by hundreds of thousands of micro-fibrils called setae, which further branch into hundreds of nanoscale spatula-like structures \cite{Gorb_book_2001,Ruibal1965,Niederegger2006}. These spatulae adhere to the substrate using intermolecular van der Waals forces \cite{Hiller1969,Autumn2000,Autumn2002b}. In their pioneering work, Autumn et al. \cite{Autumn2000} measured that a single seta of a Tokay Gecko, with proper orientation, perpendicular preloading, and a parallel drag on the substrate, generates frictional forces as high as approximately $200\,\upmu$N. 
			
			In the last two decades, researchers have extensively studied the structure, function, properties, and applications of the gecko adhesive system \cite{Russell2002,Autumn2007,Pugno2008a,Asbeck2009,Hu2012,Drotlef2017, Hou2018}. Sitti and Fearing \cite{Sitti2003} modeled the spatuale as cantilever beams to study the effect of length, diameter, stiffness, and density. Gao et al. \cite{Gao2005} used finite elements with cohesive zone models to analyse the adhesion of a gecko seta. They observed that the seta detaches when the angle between the applied tensile load and the substrate becomes more than $30^\circ$. Huber et al. \cite{Huber2005a} used atomic force microscopy to measure the adhesive forces for a single seta with only four spatulae. Their load-drag-pull experiments revealed that when the spatulae are pulled vertically the maximum pull-off force for a single spatula is close to $10$\,nN. 
			
			Tian et al. \cite{Tian2006} proposed a theoretical model for estimating the adhesion and friction forces between a single spatula and a rigid substrate. However, they did not consider how the frictional forces vary within the peel zone. They also did not consider the elastic loading and unloading of the spatula as it is being pulled. Pesika et al. \cite{Pesika2007} proposed a peel-zone model similar to Kendall's peeling model \cite{Kendall1975} but differing in an angle-dependent multiplier. Chen et al. \cite{Chen2009} used the Kendall peeling model and finite element analysis to study the effect of pre-tension on the spatula peeling characteristics such as peeling force, peeling angle, and critical detachment angle. Begley et al. \cite{Begley2013} presented an analytical mode that accounts for the frictional sliding of single and double sided peeling of an elastic tape. They found that the critical force required for peeling is higher for the case of frictional sliding than for pure sticking. Peng and Chen \cite{Peng2015} used an analytical model derived from the principle of minimum potential energy to examine the effect of bending stiffness on the peeling force. Labonte and Federle \cite{Labonte2016} studied the coupling of adhesion and friction in insects and showed that there is a linear relationship between friction and adhesion for large friction forces. The authors proposed that the large peeling forces at small peeling angles could be due to pre-tension arising from partial sliding close to the peeling front during detachment of the adhesive pads. Kim and Varenberg \cite{Kim2017} studied how the pulling angle influences the normal and tangential components of the pull-off force for wall-shaped adhesive microstructures using Kendall's peeling model \cite{Kendall1975}. Peng et al. \cite{Peng2019} studied the peeling behaviour of finite length thin-films using both theoretical and finite element models. They observed that unlike classical peeling models like Kendall's, which consider infinite film length, finite film length models lead to steady-state peeling process depending on the initial adhesion length. A detailed review of many adhesion and friction models of geckos along with advancements in gecko inspired synthetic adhesives can be found in the review papers by Kwak and Kim \cite{Kwak2010}, Jagota and Hui \cite{Jagota2011}, Zhou et al. \cite{Zhou2013}, Kasar et al. \cite{Kasar2018}, Seale et al. \cite{Seale2018}, Russell et al. \cite{Russell2019}, and Labonte and Federle \cite{Labonte2019}.
	
			Analytical models, based on simplified seta/spatula geometries combined with linear deformation behaviour offer limited insight into the details of seta and spatula adhesion, deformation, and stresses \cite{Sauer2009}. Therefore, Sauer \cite{Sauer2009b,Sauer2010} presented a nonlinear multi-scale frame work for studying seta and spatula adhesion based on the coarse-graining of van der Waals interaction \cite{RogerLi2007}. The multi-scale model was improved by Sauer and Holl \cite{Sauer2013}, by accounting for the detailed 3-D spatula geometry. Using a geometrically exact finite beam model \cite{Sauer2014}, Sauer \cite{Sauer2011} studied the effect of bending stiffness on spatula adhesion. Peng et al. \cite{Peng2010} employed a cohesive zone model to study the influence of different parameters on the peeling behaviour of a spatula. Cheng et al. \cite{Cheng2012} examined how a non-uniform pre-tension is generated when the spatula slides on the substrate during attachment. Gautam and Sauer \cite{Gautam2014} analysed gecko adhesion using a dynamic FE model. Using the geometrically exact beam formulation of \cite{Sauer2014}, Mergel et al. \cite{Mergel2014a} showed that the optimal shape of a peeling strip is similar to that of a gecko spatula. Mergel and Sauer \cite{Mergel2014b} extended the study to incorporate tangential contact. A detailed review of different computational methods used for solving adhesive contact is presented in \cite{Sauer2016}.
			
			The frictional behaviour of gecko setae is unlike other materials. When the setae are dragged proximally (towards the animal) along their natural curvature it results in tensile loading of the seta and yet under these tensile loads they display strong static and kinematic friction that disagrees with the Amontons' law \cite{Autumn2006}. In many of the analytical and computational models of gecko adhesion, the effect of friction is neglected. But it is known from experiments \cite{Autumn2000} that at small angles, the large adhesion forces that the gecko generates are determined mainly by frictional forces. Friction in general, emerges from different mechanisms at the molecular level. Major mechanisms that contribute to friction are: interlocking, wear, plastic deformation, adhesion, asperity deformation, viscous dissipation, and fracture \cite{Nosonovsky2007}. Therefore, a number of studies attempted to model friction due to adhesion. In the current work, the focus is on coupling of adhesion and friction, which is a major factor in gecko adhesion \cite{Autumn2006}. 
			
			Derjaguin \cite{Derjaguin1934} generalized Amontons' law of friction to include an additional parameter to account for adhesion. Majidi et al. \cite{Majidi2006} proposed a friction model based on this generalized Amontons' law for microfiber arrays. In their description, the frictional force is related to the normal load, through an effective friction coefficient $\mu^{}_{\text{eff}}$ that is a function of the sliding friction coefficient between the bodies and the interfacial shear strength per unit area $\tau$. Gravish et al. \cite{Gravish2010} studied the effect of sliding velocity on setal adhesion. They found that the setae became more sticky with increasing velocity. Puthoff et al. \cite{Puthoff2013} studied the effect of material properties and atmospheric humidity on dynamic friction characteristics of natural and synthetic gecko setal arrays. Classical van der Waals interaction can be described by the Lennard-Jones potential and this does not contain any frictional contribution. Jiang and Park \cite{Jiang2015} used an additional ``frictional potential" to describe the friction properties of layered materials. Many researchers have used cohesive zone models \cite{Cocou2010,Delpiero2010,Snozzi2013} with friction according to Coulomb's law to analyze sliding friction. Recently, Mergel et al. \cite{Mergel2018} have developed two new continuum contact models for friction due to adhesion that can capture sliding friction even under tensile loads. The first model, ``Model DI", assumes that the sliding traction during adhesion is independent of the normal distance and is equal to a constant frictional shear strength, which is related to the maximum adhesive traction. In their second model, ``Model EA", the sliding traction varies with the normal distance and is dependent on the normal traction.
			
			From the literature presented, which is by no means is an exhaustive list, it is clear that there are considerable efforts dedicated to modeling, understanding, and explaining coupled adhesion and friction in geckos. However, existing models have many shortcomings due to the intrinsic complexity of the gecko's hierarchical adhesion system. One of the aspects of gecko adhesion that has not yet been fully explored is the coupling of adhesion and friction at the spatula level. As such, the present work aims to model and understand coupled adhesion and friction in gecko spatula peeling. The peeling behaviour of a gecko spatula is studied here through a computational model. To the best of our knowledge, there has been no earlier detailed computational study in the literature that explores the coupling of adhesion and friction in gecko adhesion. For the present study, a combination of the adhesion model of Sauer and Li \cite{RogerLi2007} and the friction model ``Model EA" proposed by Mergel et al. \cite{Mergel2018} is used. A two-dimensional strip is considered to represent the gecko spatula as done by many other researchers \cite{Tian2006,Chen2009,Peng2010}. The peeling of the strip is simulated within a nonlinear finite element framework.
			
			\section{Mathematical Formulation} \label{MathForm}
			In this section, the computational formulation for adhesive frictional contact between the spatula and a flat rigid substrate is presented. The coarse grained contact model (CGCM) of \cite{RogerLi2007} is used. The contribution of tangential friction due to normal adhesion forces is derived from one of the coupled adhesion and friction models viz.``Model EA" of \cite{Mergel2018}. 
			
			\subsection{Weak Form} \label{sec:weak_form}
			The weak form governing the interaction of the deformable spatula with the rigid substrate can be written as		
			\begin{equation}
			\int_{\Omega}\text{grad}(\delta\boldsymbol{\varphi}) : \boldsymbol{\sigma} ~\mathrm{d}{v} 
			- \int_{\partial_c\Omega}\delta\boldsymbol{\varphi} \cdot \boldsymbol{t}_{\mathrm{c}}~\mathrm{d}a - \delta\Pi_{\text{ext}} = \,0\,,   
			\quad \forall\, \delta\boldsymbol{\varphi} \in \,\mathcal{V}, 
			\label{eq:WeakForm}
			\end{equation}
			where $ \mathcal{V} $ represents the space of kinematically admissible variations of deformation field $\boldsymbol{\varphi}$, and $ \delta\Pi_{\mathrm{ext}} $ denotes the virtual work of the external forces acting on the spatula. $\boldsymbol{\varphi}$ is defined as the motion mapping any arbitrary point $\boldsymbol{X}$ in the reference configuration $\Omega_{0}$ of the spatula to the deformed current location $\boldsymbol{x} \in \Omega$ and is given by $\boldsymbol{x} = \boldsymbol{\varphi}(\boldsymbol{X})$. 
			
			The first term in Eq. (\ref{eq:WeakForm}) represents the internal virtual work of the Cauchy stress tensor $ \boldsymbol{\sigma}$. For elastic material behaviour, this can be obtained from the variation of the internal energy $\Pi_\mathrm{int}$ \cite{Sauer2013}
			\begin{equation}
			\Pi_\mathrm{int} = \int_{\Omega_0}^{} W(\boldsymbol{F})\,\mathrm{d}V\,. 
			\label{eq:strain_energy}
			\end{equation} 
			
			In this study, a nonlinear Neo-Hookean material is considered, for which the strain energy density function $W$ is given by 
			\begin{equation}
			W(\boldsymbol{F}) = \frac{\Lambda}{2}\left(\ln J\right)^2 \, + \, \frac{\mu}{2}\left(\mathrm{tr}(\boldsymbol{F}\boldsymbol{F}^\mathrm{T})-3\right)\, - \, \mu\ln J,
			\end{equation}
			where $\boldsymbol{F} = \mathrm{Grad}(\boldsymbol{\varphi})$ denotes the deformation gradient and $J = \mathrm{det}(\boldsymbol{F})$ the volume change. $\Lambda$ and $\mu$ are the Lam\'{e} constants and are given as \cite{bonet2008} 
			\begin{equation}
			\mu = \dfrac{E}{2\,(1+\nu)} \,,\quad \quad \Lambda = \dfrac{2\,\mu\, \nu}{(1-2\nu)}\,.
			\label{eq:Lame} 
			\end{equation}
			
			The contact traction $\boldsymbol{t}_{\mathrm{c}}$ in the second term of Eq.~(\ref{eq:WeakForm}) is equal to the sum of the tractions due to adhesion, $\boldsymbol{t}_\mathrm{a}$, and friction, $\boldsymbol{t}_\mathrm{f}$, i.e.
			\begin{equation}
			\boldsymbol{t}_{\mathrm{c}} = \boldsymbol{t}_\mathrm{a}\, + \;\boldsymbol{t}_\mathrm{f}\,.
			\end{equation}
			
			Sauer and Li \cite{RogerLi2007} defined the following two material parameters that govern weak form~(\ref{eq:WeakForm}),
			\begin{equation}
			\label{eq:gam_parameters}
			\gamma^{}_\mathrm{W} = \dfrac{E}{\left(\dfrac{A_\mathrm{H}}{2\pi^2\,r_0^3}\right)}, \quad \gamma^{}_\mathrm{L} = \dfrac{R_0}{r_0}\,,
			\end{equation}
			where $E$ and $R_0$ represent an energy density (or stiffness) and a global length scale chosen for the system, respectively. Here, ${{A}_\mathrm{H}}$ denotes Hamaker's constant, and $r_0$ is the equilibrium distance of the original Lennard-Jones potential \cite{Israelachvili_book}
			\begin{equation}\label{eq:Lennard_Jones}
			\phi(r_\mathrm{d})  := \epsilon \left[ \left(\frac{{r}_0}{{r}_\mathrm{d}} \right)^{12} - \;\; 2\left(\frac{{r}_0}{{r}_\mathrm{d}} \right)^6 \right]\,,
			\end{equation} 
			where $\epsilon$ is an energy scale and $r_\mathrm{d}$ is the distance between two points in the interacting bodies.
			
			The second parameter in Eq.~(\ref{eq:gam_parameters}), $\gamma^{}_\mathrm{L}$, is the ratio of the global length scale $R_0$ used to normalize the geometry and the local length scale $r_0$. From the definitions, it can be seen that, $\gamma^{}_\mathrm{W}$ characterises the material stiffness in relation to the strength of adhesion, whereas $\gamma^{}_\mathrm{L}$ characterises the overall size of the geometry with respect to the range of adhesion.
			
			It should be noted that $\gamma^{}_\mathrm{W}$ and $\gamma^{}_\mathrm{L}$ are related to the dimensionless Tabor parameter \cite{Tabor1977} 
			\begin{equation}
			\mu^{}_\mathrm{Tabor} = \left(\frac{\sqrt[3]{15}\,\pi}{8}\right)^{2/3} \left(\dfrac{\gamma^{}_\mathrm{L}}{({\gamma^{}_\mathrm{W}})^2}\right)^{1/3}.
			\end{equation}

			\subsection{Adhesive and Frictional Tractions} \label{adh_fric_form}
			The adhesive contact traction $\boldsymbol{t}_\mathrm{a}$ is obtained from the variation of the adhesive contact potential as \cite{Sauer2009}
			\begin{equation}
			\label{eq:Adhesion_Traction}
			\boldsymbol{t}_\mathrm{a} =  t_\mathrm{a}\, \boldsymbol{n}_\mathrm{p}\,, \quad \quad t_\mathrm{a} = \frac{{A}_\mathrm{H}}{2\pi r_0^3 J_1 J_2}\,\left[  \frac{f_1}{45} \left(\frac{r_0}{r_\mathrm{s}}\right)^9 - \;\; \frac{f_2}{3} \left(\frac{r_0}{r_\mathrm{s}}\right)^3  \right]\,,
			\end{equation}
			where $ f_1 $ and $ f_2 $ are curvature dependent parameters of the interacting surfaces and $r_\mathrm{s}$ denotes the distance between the spatula and the substrate. Further, $J_1$ and $J_2$ represent the volume change of the spatula and the substrate at the surface, respectively. In the current work, as the substrate is considered rigid, $J_2 = 1$. Further, $\boldsymbol{n}_\mathrm{p}$ denotes the surface orientation of the substrate and is constant for flat surfaces. The adhesive traction can also be written with respect to the reference configuration, giving
			\begin{equation}
			\label{eq:Adhesive_Traction_Ref}
			\boldsymbol{T}_\mathrm{a} = T_\mathrm{a}\, \boldsymbol{n}_\mathrm{p}\,, \quad \quad T_\mathrm{a} = \frac{{A}_\mathrm{H}}{2\pi r_0^3J_2}\,\left[  \frac{f_1}{45} \left(\frac{r_0}{r_\mathrm{s}}\right)^9 - \;\; \frac{f_2}{3} \left(\frac{r_0}{r_\mathrm{s}}\right)^3  \right]\,.
			\end{equation}

			To avoid ill-conditioning, the adhesive traction is regularised for very small normal gaps $r^{}_\mathrm{s} \longrightarrow 0$. This is done by the linear extrapolation \cite{Mergel2018}
			\begin{equation}
			T_\mathrm{a} (r^{}_\mathrm{s}) = \begin{cases} 
			T_\mathrm{a}(r^{}_\mathrm{s}),  &  \quad r^{}_\mathrm{s} \geq r^{}_\mathrm{reg}, \\
			T_\mathrm{a} (r^{}_\mathrm{reg}) + T_\mathrm{a}^\prime(r^{}_\mathrm{reg}) (r^{}_\mathrm{s} - r^{}_\mathrm{reg}) & \quad r^{}_\mathrm{s} < r^{}_\mathrm{reg},
			\end{cases} 
			\end{equation}
			where $r_\mathrm{reg}$ is a regularisation distance and is chosen to be equal to equilibrium distance $r_\mathrm{eq}$ \cite{Mergel2018}. $T_\mathrm{a}^\prime$ indicates the derivative of the adhesive traction $T_\mathrm{a}$ with respect to $r^{}_\mathrm{s}$ and is given~as
			\begin{equation}
			T_\mathrm{a}^\prime (r_\mathrm{s}) = \dfrac{\partial T_\mathrm{a}(r_\mathrm{s})}{\partial r_\mathrm{s}} =  - \dfrac{A_\mathrm{H}}{2\pi r_0^4} \left[\dfrac{1}{5}\left(\dfrac{r_0}{r_\mathrm{s}}\right)^{10} \, - \, \left(\dfrac{r_0}{r_\mathrm{s}}\right)^3 \right]\,.
			\end{equation}
			
\begin{figure}[h!]
				\begin{center}
					\includegraphics[scale=0.29]{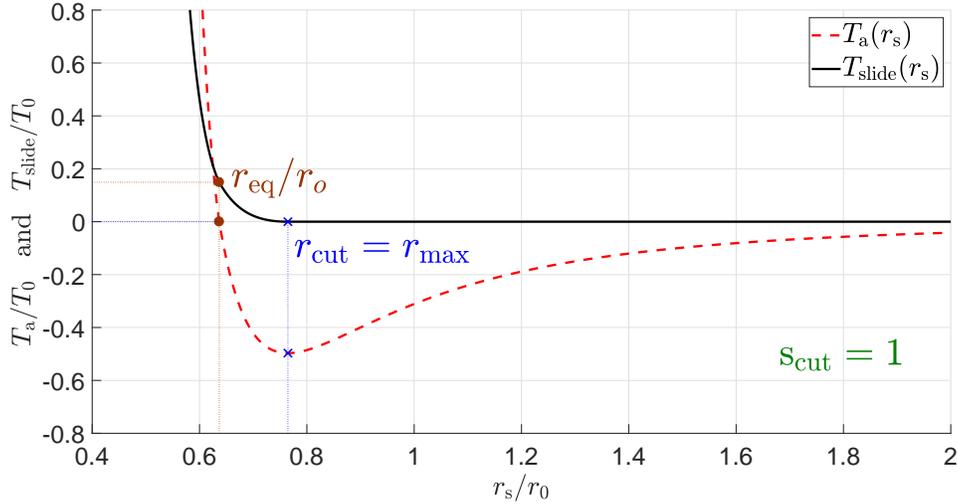}
					\caption{Variation of adhesive traction $T_\mathrm{a}$ and sliding threshold $T_\mathrm{slide}$ with the normal distance $r_\mathrm{s}$ according to adhesive friction model ``EA" of Mergel et al. \cite{Mergel2018}. Here, $T_0 = A_\mathrm{H}/(2\pi r_0^3)$, $\mu_\mathrm{s} =0.3$, and $s_\mathrm{cut} = 1$.}
					\label{fig:Adh_Tslide}
				\end{center}
			\end{figure}

			To obtain the contribution of friction due to normal adhesive forces, ``Model EA" of \cite{Mergel2018}, is used. If the magnitude of the tangential traction $T_\mathrm{f} = \left\Vert\boldsymbol{T}_\mathrm{f}\right\Vert$ exceeds a certain sliding threshold $T_{\mathrm{slide}}$, the surfaces no longer stick together and start sliding. Before defining the sliding threshold, a cut-off distance $r_\mathrm{cut}$, which is the distance up to which the frictional forces are active between the interacting surfaces, is defined as
			\begin{equation}
			r_\mathrm{cut} = s_\mathrm{cut}\,r_{\text{max}} + (1-s_\mathrm{cut})\,r_\mathrm{eq}, \quad \quad s_\mathrm{cut} \; \in \; [0,1]\,,
			\end{equation}
			where $r_\mathrm{max}$ is the distance at which the adhesive traction $T_\mathrm{a}$ given by Eq.~(\ref{eq:Adhesive_Traction_Ref}) reaches the global minimum (maximum attraction) and $r_\mathrm{eq}$ is the equilibrium distance where the adhesive traction becomes zero, see Figure~\ref{fig:Adh_Tslide}. Then, the sliding threshold is defined as (see Figure~\ref{fig:Adh_Tslide}),
			\begin{equation}
			T_\mathrm{slide}(r_\mathrm{s}) = \begin{cases} 
			\displaystyle\frac{\mu^{}_{\mathrm{s}}}{J_{\text{cl}}}\left[T_\mathrm{a}(r_\mathrm{s}) - T_\mathrm{a}(r_\mathrm{cut})\right], &  \quad r_\mathrm{s} < r_{\text{cut}}, \\
			\displaystyle \quad \quad \quad 0, & \quad r_\mathrm{s} \geq r_{\text{cut}},
			\end{cases} 
			\label{eq:Sliding_Threshold}
			\end{equation}
			where $J_{\text{cl}}$ is the local surface stretch of the neighbouring body ($=1$ if the neighbouring body is rigid). For biological adhesives, the static friction coefficient takes similar values as the kinetic friction coefficient \cite{Mergel2018}. As such, in this work, it is assumed that the value of the friction coefficient $\mu^{}_\mathrm{s}$ is the same for sticking and sliding. 
						
			Then, the tangential contact traction $\boldsymbol{T}_\mathrm{f} (\boldsymbol{r}_\mathrm{s}, \boldsymbol{r}_\mathrm{T})$ is a function of both normal gap $\boldsymbol{r}_\mathrm{s}$ and tangential slip $\boldsymbol{r}_\mathrm{T}$ and depending on whether the bodies are sticking or slipping, it satisfies the following condition
			\begin{equation}
			\label{eq:friction_traction}
			\left\Vert\boldsymbol{T}_\mathrm{f} (\boldsymbol{r}_\mathrm{s}, \boldsymbol{r}^{}_\mathrm{T})\right\Vert\begin{cases} 
			= T_\mathrm{slide}  &  \quad \text{for sliding,} \\
			< T_\mathrm{slide} & \quad \text{for sticking.}
			\end{cases} 
			\end{equation}
			
			Frictional laws are often formulated in analogy to constitutive equations describing elasto-plasticity in continuum mechanics \cite{wriggers_contact_mechanics_book}. 
			Therefore, the tangential gap is assumed to consist of a sticking part $\boldsymbol{r}_\mathrm{T}^e$, corresponding to the elastic deformation and a sliding part $\boldsymbol{r}_\mathrm{T}^s$, corresponding to inelastic or plastic deformation, i.e.,
			\begin{equation}
			\boldsymbol{r}_\mathrm{T} = \boldsymbol{r}_\mathrm{T}^e + \boldsymbol{r}_\mathrm{T}^s\,.
			\end{equation}
			
			A linear force-gap dependence during sticking is used to calculate the tangential traction, which corresponds to a  penalty regularisation of the sticking constraint as
			\begin{equation}
			\boldsymbol{T}_\mathrm{f} = -\varepsilon_\mathrm{T} \boldsymbol{r}_\mathrm{T}^e\,,
			\end{equation}
			where the $\varepsilon_\mathrm{T}$ is equivalent to the elasticity modulus. To determine the gap $\boldsymbol{r}_\mathrm{T}^s$ due to sliding, an evolution equation is required, which is derived from the principle of maximum dissipation. The dissipation due to the plastic slip $\boldsymbol{r}_\mathrm{T}^s$ and the relative plastic tangential velocity $\dot{\boldsymbol{r}}_\mathrm{T}^s$ is given as 
			\begin{equation}
			\mathcal{D}_s = -\boldsymbol{T}_\mathrm{f} \cdot \dot{\boldsymbol{r}}_\mathrm{T}^s ~\geq~ 0 \,.
			\end{equation}
			
			Next, consider a domain of feasible contact tractions
			\begin{equation}
			\mathbb{E}_t := \{\boldsymbol{T}_\mathrm{f} \in \mathbb{R}^2 \,\,|\,\, f_s(\boldsymbol{T}_\mathrm{f})\leq 0\}\,,
			\end{equation}
			where the set $f_s$ contains all the possible tangential tractions $\boldsymbol{T}_\mathrm{f}^*$ that satisfy, 
			\begin{equation}
			f_s = \left\Vert\boldsymbol{T}_\mathrm{f}^*\right\Vert - \left\Vert\boldsymbol{T}_\mathrm{slide}\right\Vert \leq 0\,.
			\label{eq:friction_ineq}
			\end{equation}
			
			According to the principle of maximum dissipation: for the given inelastic/plastic slip velocity $\dot{\boldsymbol{r}}_\mathrm{T}^s$, the true tangential traction $\boldsymbol{T}_\mathrm{f}$ resisting the sliding motion is such that it maximizes the dissipation $\mathcal{D}_\mathrm{s}$ \cite{wriggers_contact_mechanics_book} i.e.,
			\begin{equation}
			\left(\boldsymbol{T}_\mathrm{f} - \boldsymbol{T}_\mathrm{f}^*\right)\cdot\, \dot{\boldsymbol{r}}_\mathrm{T}^s \geq 0 \,\;\; \forall\; \boldsymbol{T}_\mathrm{f}^* \in \mathbb{E}_t \,,
			\end{equation}
			which leads to the constitutive evolution equation for the plastic slip,
			\begin{equation}
			\dot{\boldsymbol{r}}_\mathrm{T}^s = \gamma \boldsymbol{n}_T \quad \quad \text{with} \quad \boldsymbol{n}_\mathrm{T} = \frac{\boldsymbol{T}_\mathrm{f}}{\left\Vert\boldsymbol{T}_\mathrm{f}\right\Vert} \,,
			\label{eq:evo_equation}
			\end{equation}
			where the parameter $\gamma\geq0$ is computed from the Karush-Kuhn-Tucker conditions of plasticity given by
			\begin{equation}
			\gamma~\geq~0, \quad \quad f_s(\boldsymbol{T}_\mathrm{f})~\leq~0, \quad \quad \gamma \cdot f_s(\boldsymbol{T}_\mathrm{f}) = 0 \,.
			\end{equation} 
			
			For solving Eq.~(\ref{eq:friction_ineq}) and Eq.~(\ref{eq:evo_equation}), a predictor-corrector algorithm is used, see Algorithm~1.  The approach is similar to classic friction algorithms	 \cite{wriggers_contact_mechanics_book}.
			\\
			\noindent\rule{17.6cm}{1pt}\\
			\textbf{Algorithm 1: Predictor-Corrector algorithm for adhesive friction with rigid~substrates.}\\
			\text{\noindent\rule{17.6cm}{1pt}}
			\begin{enumerate}
				\item $\text{Known values at pseudo-time}$ $t$, $t+\Delta t$: $\quad \boldsymbol{r}_\mathrm{T}^{t}, \quad \boldsymbol{r}_\mathrm{T}^{t+\Delta t}, \quad (\boldsymbol{T}_\mathrm{a})^{t+\Delta t}$ 
				\item Compute trial step (Predictor):$\quad \quad (\boldsymbol{T}_\mathrm{f})^{t + \Delta t}_\mathrm{tr}\, = \,-\varepsilon_\mathrm{T}\, \left[(\boldsymbol{r}_\mathrm{T})^{t+\Delta t} \,-\, (\boldsymbol{r}_\mathrm{T}^s)^t\right]$
				\item[]                                    $\hspace{6cm} \quad (f_s)^{t+\Delta  t}_\mathrm{tr} \,=\, \left\Vert(\boldsymbol{T}_\mathrm{f})^{t + \Delta t}_\mathrm{tr}\right\Vert \,- \,\left\Vert(\boldsymbol{T}_\mathrm{slide})^{t+\Delta t}\right\Vert$ 
				\item Check: $\quad\bf{if}$ $(f_s)^{t+\Delta t}_\mathrm{tr} \leq 0$ $\bf{then}$
				\item[] $\hspace{2cm} \bf{sticking}$; set ${\Delta \gamma}^{t+\Delta t} \,=\, \dfrac{\left\Vert(\boldsymbol{T}_\mathrm{f})^{t + \Delta t}_\mathrm{tr}\right\Vert\, -\, \left\Vert(\boldsymbol{T}_\mathrm{slide})^{t+\Delta t}\right\Vert}{-\varepsilon_\mathrm{T}} \,=\, 0 $;
				\item[]     $\hspace{2cm} \quad$ go to step 5. 
				\item[]     $\hspace{1.2cm}\quad$ $\bf{elseif}$ $(f_s)^{t+\Delta t}_\mathrm{tr} > 0$ $\bf{then}$ 
				\item[]     $\hspace{2cm} \quad \bf{sliding}$ ; go to step 4.
				\item Radial return mapping (Corrector):
				\item[]   $\hspace{2cm}\quad \bf{Solve}$ $(f_s)^{t+\Delta t}_\mathrm{tr} \,= 0$  \; $\bf{for}$ \;${\Delta \dot{\gamma}}^{t+\Delta t}$ 
				\item[] $\hspace{2cm} \quad$ go to step 5.
				\item[]   $\hspace{2cm} \quad {\Delta \gamma}^{t+\Delta t} \;=\; \dfrac{\left\Vert(\boldsymbol{T}_\mathrm{f})^{t + \Delta t}_\mathrm{tr}\right\Vert - \left\Vert(\boldsymbol{T}_\mathrm{slide})^{t+\Delta t}\right\Vert}{-\varepsilon_\mathrm{T}}$
				\item Update:
				$\quad (\boldsymbol{T}_\mathrm{f})^{t + \Delta t} \;=\; (\boldsymbol{T}_\mathrm{f})^{t + \Delta t}_\mathrm{tr} - {\Delta \gamma}^{t+\Delta t}\, \epsilon_\mathrm{T}\, \boldsymbol{n}_\mathrm{T}^{t+\Delta t}$ 
				\item[] $\hspace{2cm}\text{where} ~~~\boldsymbol{n}_\mathrm{T}^{t + \Delta t} \,= \,\dfrac{(\boldsymbol{T}_\mathrm{f})^{t + \Delta t}_\mathrm{tr}}{\left\Vert(\boldsymbol{T}_\mathrm{f})^{t + \Delta t}_\mathrm{tr}\right\Vert}$ 
				\item[]   $\hspace{2cm}\quad (\boldsymbol{r}_\mathrm{T}^s)^{t+\Delta t} \,=\, (\boldsymbol{r}_\mathrm{T}^s)^{t} \,+ \,{\Delta \gamma}^{t+\Delta t}\,\boldsymbol{n}_\mathrm{T}^{t+\Delta t}$
				\item[] $\hspace{2cm} \quad$ go to step 1. 
			\end{enumerate}
			\noindent\rule{13.6cm}{1pt}\\
			
			\section{Finite Element Formulation} \label{FE_form}
			Next, the finite element (FE) formulation used for the coupled adhesion-friction model, which is employed to analyse the peeling of a gecko spatula from a flat rigid substrate, is presented. Further, the surface enrichment strategy used in the FE discretization is also discussed. It should be noted that this is a deterministic formulation. 
			\subsection{FE discretization of the weak form}
			The weak form in Eq.~(\ref{eq:WeakForm}) is discretized using the finite element method. Therefore, the spatula is discretized into a number of volume and surface elements containing $n_{e}$ nodes ($= n_{ve}$ for a volume element and  $n_{se}$ for a surface element, respectively). In each element, the displacement field $\boldsymbol{u}$ and its variation $\boldsymbol{\varphi}$ is then approximated by the interpolations
			\begin{equation}
			\label{eq:FE_discretisation}
			\boldsymbol{u}(\boldsymbol{x}) \approx \mathbf{N}_e(\boldsymbol{x})\textbf{u}^e, \quad \delta\boldsymbol{\varphi(x)} \approx \mathbf{N}_e(\boldsymbol{x})\textbf{v}^e\,,
			\end{equation}
			where $\textbf{u}^e$ and $\textbf{v}^e$ are the nodal displacement and variations of element $e$, and the matrix $\mathbf{N}_e$ is a matrix formed by $n_{e}$ shape functions of the element and is given by
			\begin{equation}
			\mathbf{N}_e = [N_1^e\boldsymbol{I},\quad N_2^e\boldsymbol{I}, \quad ... \quad N_{n_{e}}^e\boldsymbol{I}]\,,
			\end{equation}
			where $\boldsymbol{I}$ represents the identity tensor. 
			
			Using the relations in Eq.~(\ref{eq:FE_discretisation}) and performing an assembly over all the volume and surface elements, the weak form in Eq.~(\ref{eq:WeakForm}) can then be rewritten into
			\begin{equation}
			\label{eq:EOM_A}
			\mathbf{v}^\mathrm{T}\left[\mathbf{f}_{\text{int}} + \mathbf{f}_\mathrm{c} - \mathbf{f}_{\text{ext}}\right] = \mathbf{0}, \quad \forall \; \mathbf{v} \, \in \, \mathcal{V}_k\,,
			\end{equation}
			where $\mathbf{f}_{\mathrm{int}}$, $\mathbf{f}_\mathrm{c}$, and $ \mathbf{f}_{\mathrm{ext}}$ denote the global internal, contact, and external nodal force vectors, respectively. In the current work, as there are no external forces acting on the spatula $\mathbf{f}_{\mathrm{ext}} = \mathbf{0}$. 
			
			Since, the the variations $ \mathbf{v} $ are arbitrary,  Eq.~(\ref{eq:EOM_A}) leads to
			\begin{equation}
			\label{eq:EOM}
			\mathbf{f(\mathbf{u})}:= \,\, \mathbf{f}_{\text{int}} + \mathbf{f}_\mathrm{c} - \mathbf{f}_{\mathrm{ext}} = \mathbf{0}\,.
			\end{equation}
			
			The global nodal force vectors in Eq.~(\ref{eq:EOM}) are obtained from assembling the individual elemental contributions. 
			
			The elemental internal force vector $\mathbf{f}^e_{\mathrm{int}}$ acting on element $\Omega^e$ is given as
			\begin{equation}
			\mathbf{f}_{\mathrm{int}}^e = \int\limits_{{\Omega}^e} \mathbf{B}_{e}^\mathrm{T}~\boldsymbol{\sigma} ~\mathrm{d}{v},
			\label{eq:fint}
			\end{equation}
			where $\mathbf{B}_e$ is an array that contains the derivatives of the $n_{ve}$ number of shape functions \cite{zeink_book} and $\boldsymbol{\sigma}$ is the Cauchy stress tensor given by \cite{bonet2008}
			\begin{equation}
			\label{eq:CauchyTensor}
			\boldsymbol{\sigma} ~=~ \frac{\Lambda}{J}~(\text{ln}J)~\boldsymbol{I} ~+ ~\frac{\mu}{J}~(\boldsymbol{F}\boldsymbol{F}^\mathrm{T}- \,\boldsymbol{I})\,,
			\end{equation}
			and arranged in Voigt notation to suit Eq.~(\ref{eq:fint}). Introducing a modified shear modulus: $\mu^* := \mu - \Lambda \ln J$, the spatial elasticity tensor of the Neo-Hookean material can be written as 
			\begin{equation}
			\mathbb{C} = \frac{\Lambda}{J}~\boldsymbol{I}\otimes\boldsymbol{I} ~+ ~\dfrac{2\mu^*}{J}~\,\mathbb{I}\,,
			\label{eq:elast_tensor}
			\end{equation}
			where $\mathbb{I}$ represents the fourth-order symmetric identity tensor, whose components can be expressed in terms of the Kronecker delta: $\mathbb{I}_{ijkl} = \frac{1}{2}\left( \delta_{ik}\delta_{jl} + \delta_{il}\delta_{jk} \right)$.
			
			The spatial elasticity tensor $\mathbb{C}$ can be written in Voigt notation as
			\begin{equation}
			\mathbb{C}_v = \frac{1}{J} \begin{bmatrix}
			2\mu^* + \Lambda & \Lambda & \Lambda & 0  & 0 & 0 \\
			\Lambda& 2\mu^* + \Lambda & \Lambda & 0 & 0 & 0 \\
			\Lambda & \Lambda & 2\mu^* + \Lambda & 0 & 0 & 0 \\
			0 & 0 & 0 & \mu^*  & 0 & 0 \\
			0 & 0 & 0 & 0  & \mu^* & 0 \\
			0 & 0 & 0 & 0  & 0 & \mu^*
			\end{bmatrix}
			\label{eq:elast_voigt}
			\end{equation}
			
			The elemental contact force vector acting on $\partial_c\Omega^e$ is given as
			\begin{equation}\label{eq:cont_forc_def}
			\mathbf{f}_{\mathrm{c}}^e =  -\int\limits_{\partial_c\Omega^e}\mathbf{N}_e^\mathrm{T}~\boldsymbol{t}_{\mathrm{c}}~\mathrm{d}a\,,
			\end{equation}
			which can be transformed to the reference configuration using Nanson's formula,
			\begin{equation}
			\boldsymbol{n}_\mathrm{s}\,\mathrm{d}a = J\boldsymbol{F}^\mathrm{-T}\boldsymbol{N}_\mathrm{s}\,\mathrm{d}A\,,
			\end{equation}  
			where $\mathrm{d}a$ denotes the surface element of the spatula with the orientation $\boldsymbol{n}_\mathrm{s}$ in the deformed configuration $\partial_c \Omega^e$, and $\mathrm{d}A$ denotes the corresponding surface element with the orientation $\boldsymbol{N}_\mathrm{s}$ in the undeformed reference configuration $\partial_c \Omega_0^e$. Then Eq.~(\ref{eq:cont_forc_def}) becomes,
			\begin{equation}\label{eq:cont_forc_orig}
			\mathbf{f}_{\text{c}}^e =  -\int\limits_{\partial_c\Omega_0^e}\mathbf{N}_e^\mathrm{T}~\Theta~\boldsymbol{T}_{\mathrm{c}}~\text{d}A\,,
			\end{equation}
			with 
			\begin{equation}
			\Theta := -\boldsymbol{n}_\mathrm{p} \cdot \boldsymbol{F}^{\mathrm{-T}}\boldsymbol{N}_\mathrm{s}\,,
			\end{equation}
			where $\boldsymbol{n}_\mathrm{p}$ denotes the surface orientation of the substrate and is constant for the flat, rigid substrate considered in the current work. Also, for gecko adhesion $\Theta$ can be approximated as unity as shown by Sauer and Wriggers \cite{Sauer2009}.
			
			The elemental contact force vector in Eq.~(\ref{eq:cont_forc_orig}) can then be written as the sum of the contributions of the elemental force vectors due to normal adhesive traction $\boldsymbol{T}_\mathrm{a}$ and tangential frictional traction $\boldsymbol{T}_\mathrm{f}$
			\begin{equation}\label{eq:tot_ele_forc}
			\mathbf{f}_{\text{c}}^e \,=\, -\int\limits_{\partial_c\Omega_0^e}\mathbf{N}_e^\mathrm{T}~\boldsymbol{T}_{\mathrm{a}}~\mathrm{d}A \,-\int\limits_{\partial_c\Omega_0^e}\mathbf{N}_e^\mathrm{T}~\boldsymbol{T}_{\mathrm{f}}~\mathrm{d}A\,.
			\end{equation}
			
			The nonlinear equation in Eq.~(\ref{eq:EOM}), is solved iteratively using the Newton-Raphson method. Therefore, the tangent stiffness matrix $\mathbf{k}^e_\mathrm{c}$ is formed by the elemental contributions and is given by (see \cite{Sauer2009} for a detailed derivation)
			\begin{equation}\label{eq:tang_stiff}
			\mathbf{k}^e_\mathrm{c} = \frac{\partial \mathbf{f}_{\mathrm{c}}^e}{\partial \mathbf{u}_e} = - \int_{\partial_c\Omega^e_{0}} \mathbf{N}_e^{\mathrm{T}}~\left(\dfrac{\partial \boldsymbol{T}_\mathrm{a}}{\partial \boldsymbol{x}} + \dfrac{\partial \boldsymbol{T}_\mathrm{f}}{\partial \boldsymbol{x}} \right)~\mathbf{N}_e~ \mathrm{d}A\,.
			\end{equation}
			
			The derivatives of the adhesive and frictional tractions in Eq.~(\ref{eq:tang_stiff}) are given as
			\begin{eqnarray}\label{eq:tang_stiff_der}
			\frac{\partial \boldsymbol{T}_\mathrm{a}}{\partial \boldsymbol{x}}  &=& \dfrac{\partial T_\mathrm{a}(r_\mathrm{s})}{\partial r_\mathrm{s}}\, \boldsymbol{n}_\mathrm{p} \otimes \boldsymbol{n}_\mathrm{p}\,. \\
			\dfrac{\partial \boldsymbol{T}_\mathrm{f}}{\partial \boldsymbol{x}} &= &  \begin{cases} -\varepsilon_\mathrm{T} \ \, \boldsymbol{a}_\mathrm{p} \otimes \boldsymbol{a}_\mathrm{p} &\quad \quad \text{for elastic step i.e., sticking}\,, \\
			\dfrac{\partial T_\mathrm{slide}}{\partial r_\mathrm{s}} \,\, \boldsymbol{n}_\mathrm{T} \otimes \boldsymbol{a}_\mathrm{p} &\quad \quad \text{for inelastic step i.e., sliding}  \,.
			\end{cases} 
			\end{eqnarray}
			
			\subsection {Enrichment strategy}
			As discussed by Sauer \cite{RogerEnriched2011}, in solving a strip peeling problem for the case of strong adhesion, numerical difficulties arise because of the highly nonlinear nature of the van der Waals forces. For coarse finite element meshes, the use of four-noded quadrilateral elements can lead to poor convergence in the Newton-Raphson iterations. These issues can be resolved by using a very fine mesh, which lead to undesirably high computational cost. In order to address this issue, Sauer \cite{RogerEnriched2011} introduced surface enriched FE based on p-refinement with fourth order Lagrange polynomials, denoted as Q1C4. This is employed in the current work. This means that, the contact elements of the spatula are discretized with Q1C4 finite elements whereas the bulk of the spatula domain is discretized using standard quadrilateral finite elements (denoted as Q1). 
			\begin{figure}[h!]
				\begin{center}
					\includegraphics[scale=0.4]{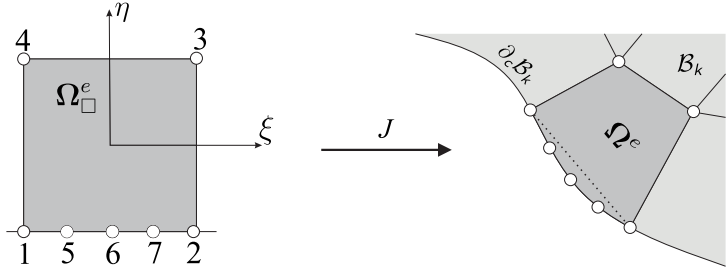}
					\caption{A schematic representation of Q1C4 enriched element and its corresponding mapping to physical space. (Adapted from Sauer \cite{RogerEnriched2011})}
					\label{fig:enrich}
				\end{center}
			\end{figure}
			
			The formulation for Q1C4 elements is developed from standard Q1 elements for which the displacement fields in an element are given as $\boldsymbol{u}_e^h = \sum_{i=1}^{4} N_i^0\mathbf{u}_i$. The shape functions $N_i^0$, for Q1 element, are given as
			\begin{eqnarray}
			N_1^0(\xi,\eta) &=&  \frac{1}{4}(1-\xi)(1-\eta)\,,  \\
			N_2^0(\xi,\eta) &=&  \frac{1}{4}(1+\xi)(1-\eta)\,,   \\
			N_3^0(\xi,\eta) &=&    \frac{1}{4}(1+\xi)(1+\eta)\,,  \\
			N_4^0(\xi,\eta) &=&   \frac{1}{4}(1-\xi)(1+\eta)\,.
			\end{eqnarray}
			
			In the formation of Q1C4, three extra nodes are inserted at $(\xi, \eta) = (0, -1), (-0.5, -1)$, and $(0.5,-1)$ as shown in Figure~\ref{fig:enrich}, and the additional shape functions are defined as 
			\begin{eqnarray}
			N_5(\xi,\eta) &=&  ~~~2\left(\xi^4 - \frac{5}{4}\xi^2 + \frac{1}{4}\right)(1-\eta)\,,  \\
			N_6(\xi,\eta) &=&  -\frac{4}{3}\left(\xi^4 - \frac{1}{2}\xi^3 - \xi^2 + \frac{1}{2}\xi\right)(1-\eta)\,,   \\
			N_7(\xi,\eta) &=&    -\frac{4}{3}\left(\xi^4 + \frac{1}{2}\xi^3 - \xi^2 - \frac{1}{2}\xi\right)(1-\eta)\,.  
			\end{eqnarray}
			
			Then, with $N_3^0$ and $N_4^0$ remaining same, the following modified shape functions are obtained for the enriched surface element
			\begin{eqnarray}
			N_1 = N_1^0 - \frac{1}{2}N_5 - \frac{3}{4}N_6 - \frac{1}{4}N_7\,, \\
			N_2 = N_2^0 - \frac{1}{2}N_5 - \frac{1}{4}N_6 - \frac{3}{4}N_7\,.
			\end{eqnarray}
			
			With this, the displacement field in the interior of the surface element is interpolated as
			\begin{equation}
			\boldsymbol{u}_e^h = \sum_{i=1}^{7}N_i\textbf{u}_i\,.
			\end{equation}
				
			\section{Spatula Model}
			
			The spatula is modelled as a thin two-dimensional strip in plain strain, similar to many studies in the literature \cite{Tian2006,Chen2009,Peng2010}. As such, the words ``strip" and ``spatula" are used interchangeably in the following text. 
			
			\subsection{Model parameters} 
\begin{figure}[h!]
				\begin{center}
					\includegraphics[scale=0.38]{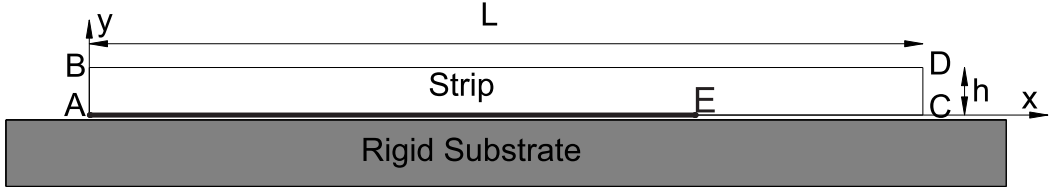}
					\caption{A deformable strip on a rigid substrate.}
					\label{fig:Strip_orig}
				\end{center}
			\end{figure}

			The dimensions of the strip are $ L\times h $ (with $L=200 R_0 $, $ h=10R_0$ where $ R_0=1 $\,nm is introduced for normalization) as shown in Figure~\ref{fig:Strip_orig} \cite{RogerEnriched2011}. The bottom $ 75\% $ of the strip surface (``AE" in Figure~\ref{fig:Strip_orig}) is assumed to be in adhesion. Hence, ``AE" is assumed to represent the pad of the spatula while ``EC" is assumed to represent the spatula shaft. The area of the gecko spatula pad is taken as, $A_\mathrm{pad} = 49,524R_0^2$ \cite{Sauer2013}. Then the average width of the spatula pad becomes $w = 330.16R_0$. 
		
			The material of the strip is taken to be an isotropic non-linear Neo-Hookean material (see Eq.~\ref{eq:strain_energy}). The default values for Young's modulus and Poisson's ratio are taken as $ E = 2 $ GPa and $ \nu = 0.2 $, respectively \cite{Tian2006,Sauer2013}. The corresponding forces are calculated using Eqs.~(\ref{eq:Adhesive_Traction_Ref}) and (\ref{eq:friction_traction}) with $ r_0=0.4 $\,nm and $ A_\mathrm{H}= 10^{-19} $\,J. The values considered for $E$, $A_\mathrm{H}$, and $r_0$ correspond to the gecko spatula material and the associated adhesion energy \cite{Tian2006,Sauer2013,Gautam2014}. This results in $\gamma^{}_\mathrm{W}=25.266$ and $\gamma^{}_\mathrm{L}=2.50$ according to Eq.~(\ref{eq:gam_parameters}). The friction coefficient is taken as $\mu_s = 0.3$ in agreement with experimental data on gecko seta friction on glass surfaces \cite{Autumn2000,Autumn2002b,Autumn2006}. 

			The strip is discretized into $ 240\times 12 $ quadrilateral finite elements along the  x and  y directions. Enriched contact finite elements, Q1C4, are used at the contact interface \cite{RogerEnriched2011} and the bulk of the domain is discretized using standard quadrilateral finite elements. This particular finite element discretization has been shown to be very accurate for peeling computations \cite{RogerEnriched2011}. Plane strain conditions are considered in all simulations.
			
			\subsection{Application of peeling}\label{sec:peeling} %
			
\begin{figure}[]
				\centering
				\begin{subfigure}{\linewidth}
					\begin{center}
						\includegraphics[scale=0.4]{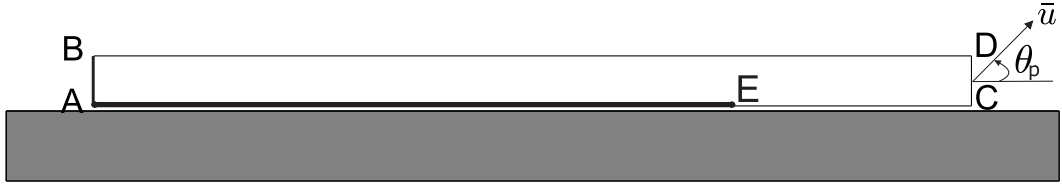}
						\caption{Peeling by applied displacement $\bar{u}$ from the undeformed configuration. The inclination of $\bar{u}$, $\theta_\mathrm{p}$, is denoted as the peeling angle.}
						\label{fig:configuration1}
					\end{center}
				\end{subfigure} \\ 
				\vspace{1.5cm}
				\begin{subfigure}{\linewidth}
					\begin{center}
						\includegraphics[scale=0.4]{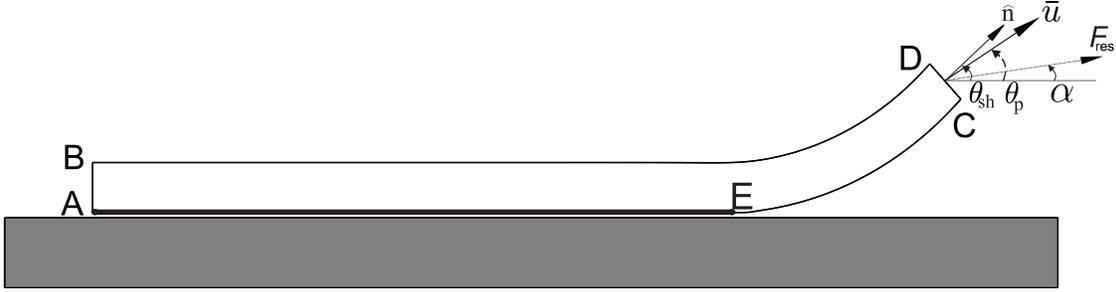}
						\caption{Peeling by applied rotation $\theta$ and then applied displacement $\bar{u}$. Here, $\theta_\mathrm{sh}$ denotes the angle between the cross-section normal $\boldsymbol{\hat{n}}$ and the substrate, while $\alpha$ denotes the angle between the resultant force $F_\mathrm{res}$ and the substrate.} 
						\label{fig:configuration2}
					\end{center}
				\end{subfigure} 
				\caption{Application of peeling.}
				\label{fig:configuration_peeling}
			\end{figure}

			Peeling can occur in two ways: \emph{tangentially-constrained} peeling and \emph{tangentially-free}  peeling. Next, each type is described.  
			\begin{enumerate}
			\item The \emph{tangentially-constrained} peeling of the strip is effected by applying a displacement $\bar{u}$ to the right end of the  strip (CD) - resting in its initial configuration - at an angle, which is denoted as the peeling angle $\theta_\mathrm{p}$ as shown in Figure~{\ref{fig:configuration1}}. 
				
			The displacement $\bar{u}$ is applied such that $u_x = \bar{u}\cos(\theta_\mathrm{p})$ and $u_y = \bar{u}\sin(\theta_\mathrm{p})$. As a result, right end CD is constrained in the tangential direction. For a spatula, this essentially translates into constraining the tangential movement of the spatula shaft. Due to this, the direction of resultant force $F_\mathrm{res}$ is different from the applied displacement $\bar{u}$.

			
			
			\item In case of \emph{tangentially-free} peeling (examined in section~{\ref{sec:DH_detachment}}), The displacement $\bar{u}$ is applied only in the y-direction i.e. perpendicular to the substrate. As such, the lateral displacement of the shaft i.e. right end CD is not constrained. Hence the resultant force $F_\mathrm{res}$ is parallel to the applied displacement $\bar{u}$.
					
			\end{enumerate}	
	
			\section{Results and Discussion}
			In this section the numerical results of the spatula peeling are presented. The normal (adhesive) and tangential (frictional) components of the resultant pull-off force $F_\mathrm{res}$ corresponding to the applied displacement $\bar{u}$ are denoted by $F_\mathrm{N}$ and $F_\mathrm{T}$, respectively. First, the numerical results for the case of \emph{tangentially-constrained} peeling are presented and the effect of various parameters -- such as the peeling angle $\theta_\mathrm{p}$, shaft angle $\theta_\mathrm{sh}$, strip thickness $h$, and material stiffness -- on the peeling behaviour is discussed in sections~{\ref{sec:desc_peel}} to {\ref{sec:Eff_bend}}. It should be noted that for this type of peeling the resultant force angle $\alpha$, which is the angle between the resultant force vector and the substrate, can be calculated as $\tan^{-1}_{}\left({F_\mathrm{N}}/{F_\mathrm{T}}\right)$ (see Figure~\ref{fig:configuration2}). The resultant force angle $\alpha$ is different from the peeling angle $\theta_\mathrm{p}$. This is due to the fact that the strip has non-zero bending stiffness and is restricted laterally as is noted in section~\ref{sec:peeling}. The results for the case of \emph{tangentially-free} peeling are presented in section~{\ref{sec:DH_detachment}}.       
			     
			\subsection{Description of the peeling curve}\label{sec:desc_peel}
			\begin{figure}[h!]
				\begin{center}
					\includegraphics[scale=0.33]{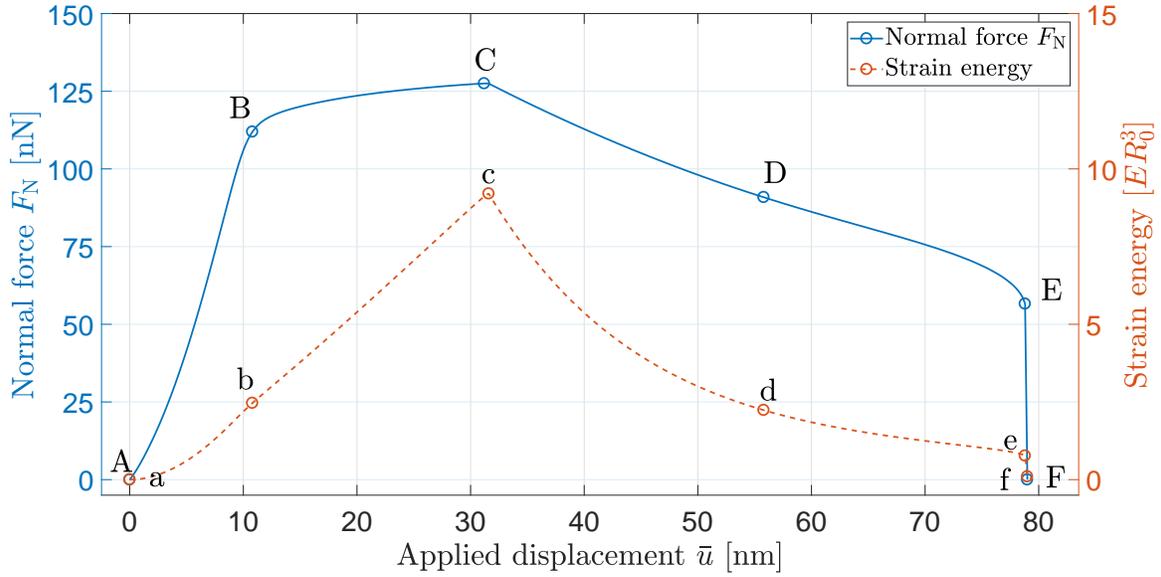} 		 
					\caption{Evolution of the normal pull-off force $F_\mathrm{N}$ and strain energy (see Eq.~({\ref{eq:strain_energy}})) with the applied displacement $\bar{u}$ for peeling angle $\theta_\mathrm{p} = 60^\circ$. Here, $ER_0^3 = 2 $ nN$\cdot$nm.}  	 \label{fig:Phase_exp}	
				\end{center}
			\end{figure}

\begin{figure}[h!]
				\centering
				\begin{subfigure}[b]{0.45\linewidth}
					\hspace{-1.15cm}
					\includegraphics[scale=0.18]{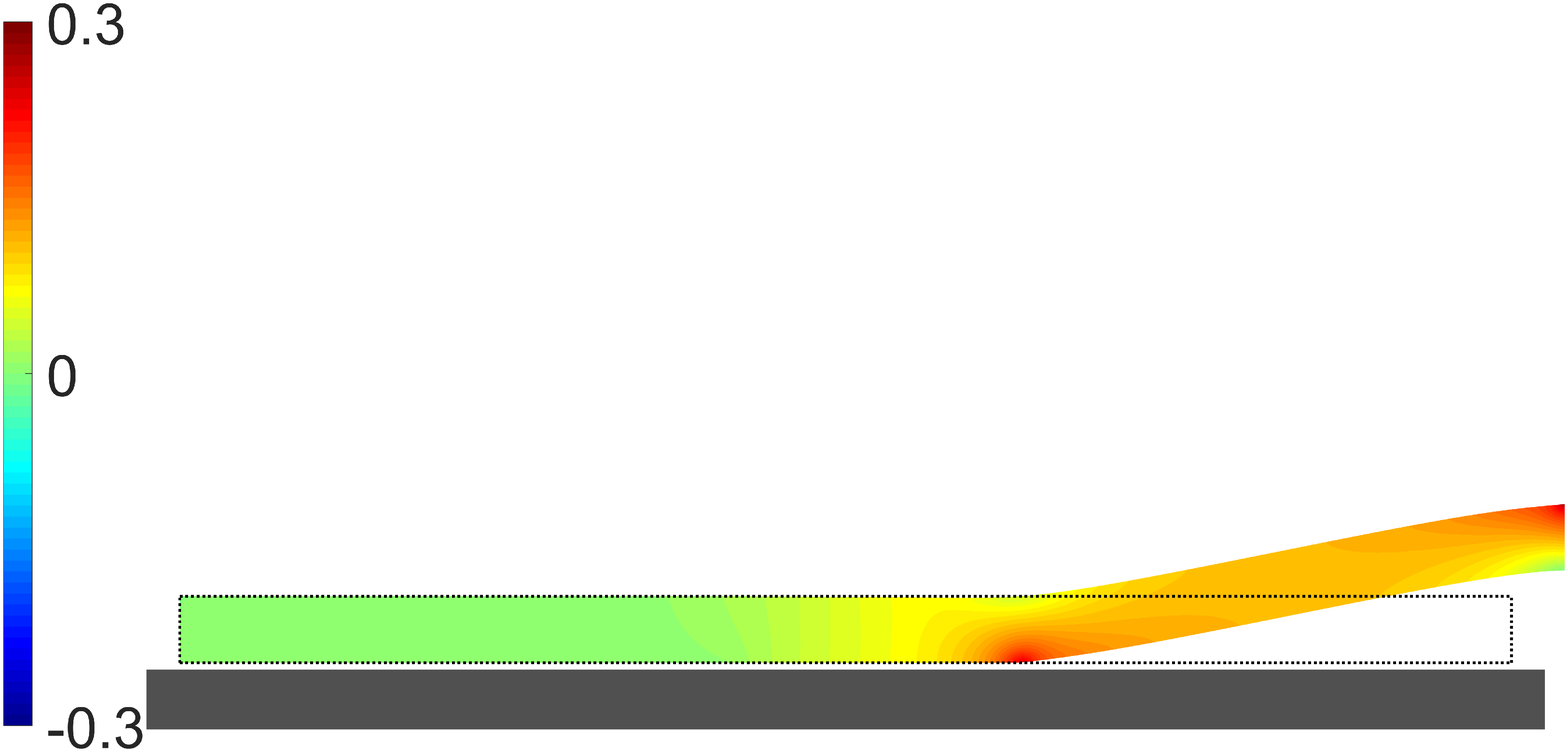}
					\caption{$\bar{u}^{}_\mathrm{B} = 16$\,nm}
					\label{fig:Deform_b}
				\end{subfigure} 
				\quad  
				\begin{subfigure}[b]{0.45\linewidth}
					\hspace{-0.65cm}
					\includegraphics[scale=0.18]{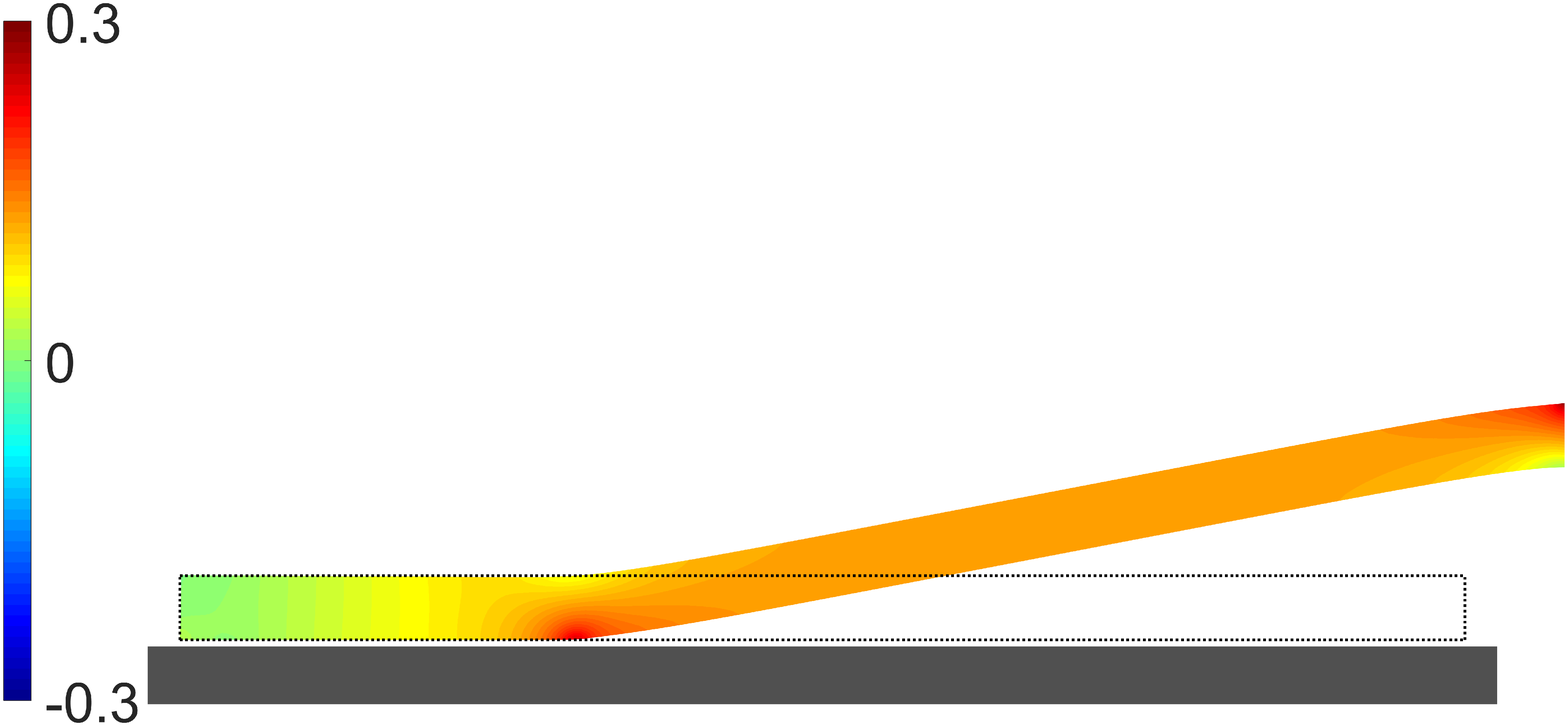}
					\caption{$\bar{u}^{}_\mathrm{C} = 31$\,nm}
					\label{fig:Deform_c}
				\end{subfigure} \\
				
				\begin{subfigure}[b]{0.45\linewidth}
					\hspace{-1.15cm}
					\includegraphics[scale=0.18]{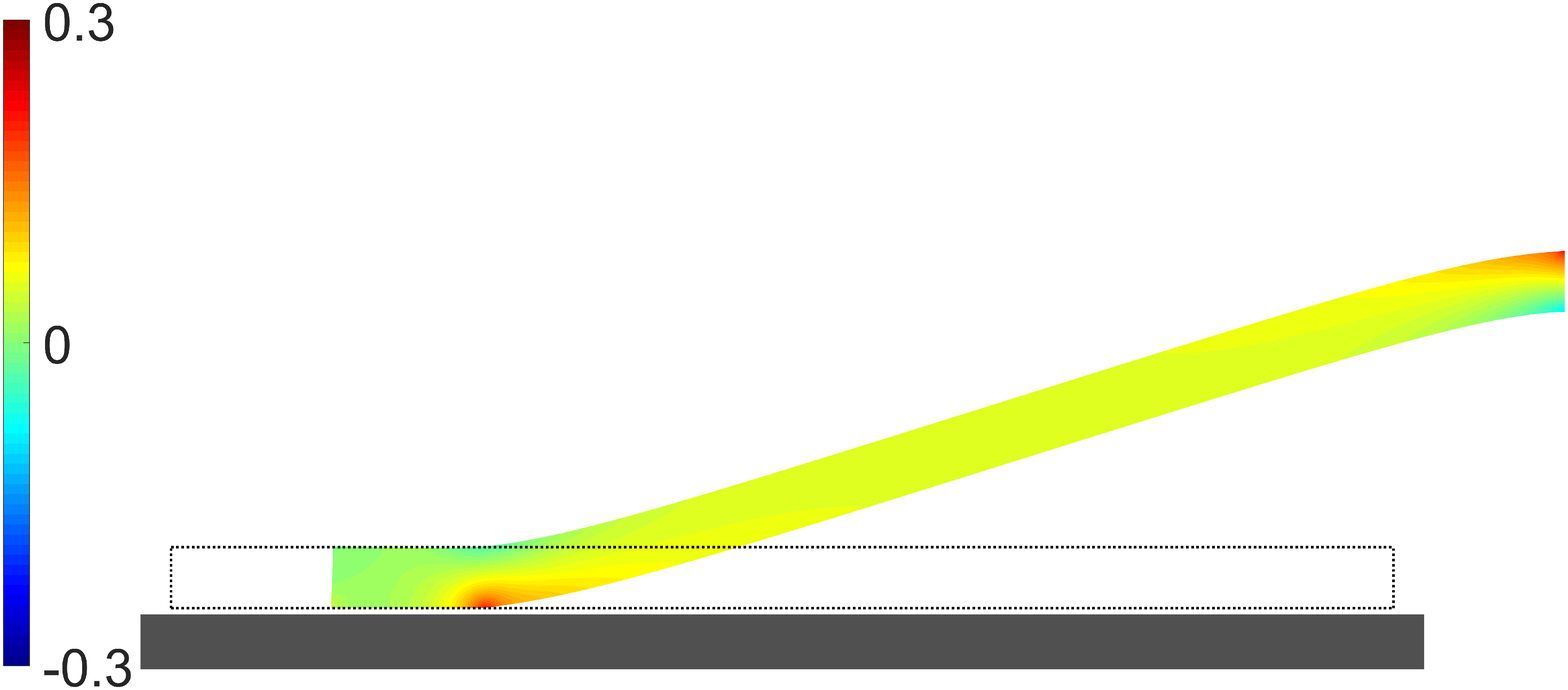}
					\caption{$\bar{u}^{}_\mathrm{D} = 56$\,nm}
					\label{fig:Deform_d}
				\end{subfigure} 
				\quad 
				\begin{subfigure}[b]{0.45\linewidth}
					\hspace{-0.65cm}
					\includegraphics[scale=0.18]{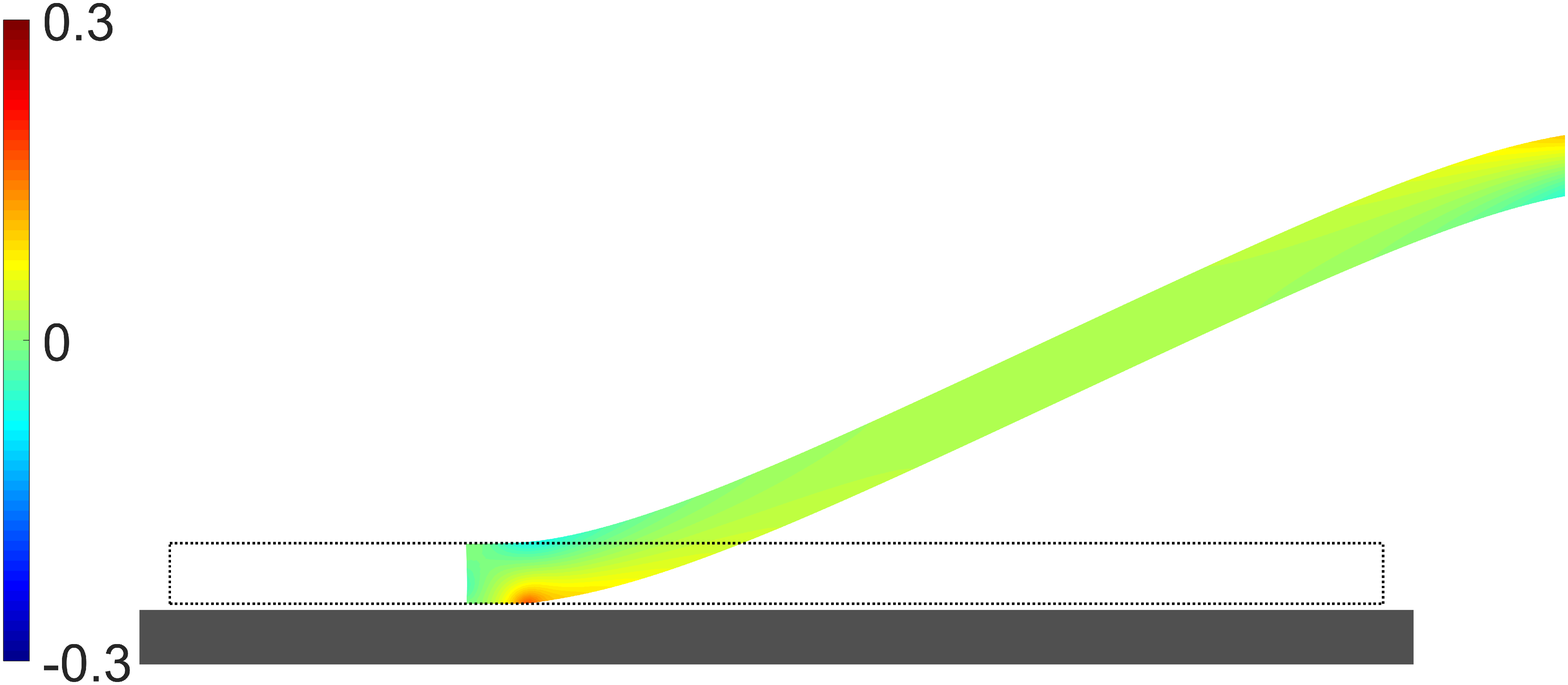}
					\caption{$\bar{u}^{}_\mathrm{E} = 79$\,nm}
					\label{fig:Deform_e}
				\end{subfigure} 
				\caption{Deformed configurations of the strip for peeling angle $\theta_\mathrm{p} = 60^\circ$ at the applied displacements $\bar{u}$ marked in Figure~\ref{fig:Phase_exp} (B to E). Up to $\bar{u}^{}_\mathrm{C}$ the spatula remains in partial sticking contact. Full sliding ensues beyond $\bar{u}^{}_\mathrm{C}$. The colorbar shows the normalised stresses $I_1/E = \text{tr}(\boldsymbol{\sigma})/E$.}
				\label{fig:Deform_60}
			\end{figure}

			The peeling of the spatula from the substrate for any given peeling angle can be divided into two phases. This can be illustrated with the help of Figure~\ref{fig:Phase_exp}, where the evolution of the normal pull-off force and the dimensionless strain energy (see Eq.~(\ref{eq:strain_energy})) with the applied displacement $\bar{u}$ for $\theta_\mathrm{p}=60^\circ$ are shown. The first phase is from the point of zero pull-off force (denoted by ``A") to the maximum value of the pull-off force (denoted by ``C"). The second phase is from point ``C" to the point at which the spatula pad snaps-off from the substrate (denoted by ``E"). The corresponding points on the strain energy curve are denoted by ``a" to ``e". Figure~\ref{fig:Deform_60} shows the strip deformation at the displacements marked in Figure~\ref{fig:Phase_exp}. 
			
			In the first phase, the spatula remains in partial sticking contact and is being continuously stretched while peeling (Figures~\ref{fig:Deform_b} and \ref{fig:Deform_c}). Moreover, in this phase, a part of the spatula pad behind the peeling front, that is still in contact, starts sliding as the spatula is pulled by the applied displacement. However, the rest of the spatula pad remains in sticking contact. This leads to stretching of the spatula, resulting in an increase in the stored strain energy of the spatula as shown in Figure~\ref{fig:Phase_exp}. As a result, this increases the force required to peel-off the spatula from the surface. After reaching the force maximum, the spatula pad is fully sliding during peeling (Figures~\ref{fig:Deform_d} and \ref{fig:Deform_e}). In this second phase, the strain energy that has been stored during the first phase, is gradually released (see Figure~\ref{fig:Phase_exp}). As the displacement is applied beyond the point ``e/E" the remaining part of the spatula pad immediately snaps-off from the surface releasing all the stored strain energy. At point ``f/F" the spatula is completely peeled off the surface.

			\subsection{Influence of the peeling angle} \label{sec:peel_ang}
			
			\begin{figure}[h!]
				\begin{center}
					\includegraphics[scale = 0.29]{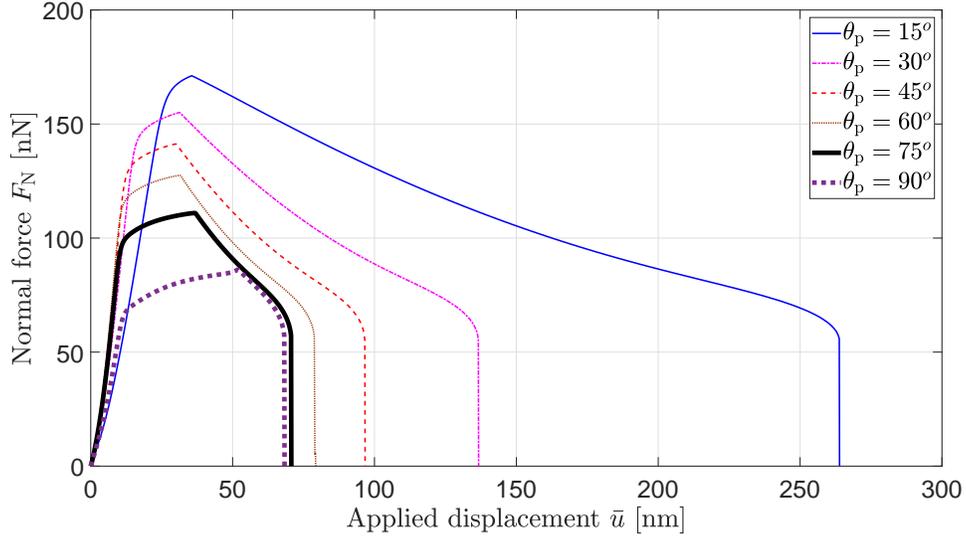} 			   
					\caption{Evolution of the normal pull-off force $F_\mathrm{N}$ with the applied displacement $\bar{u}$ for different peeling angles $\theta_\mathrm{p}$.}   	 \label{fig:Norm_Forc}	
				\end{center}
			\end{figure}
			
			\begin{figure}[h!]
				\begin{center}
					\includegraphics[scale = 0.29]{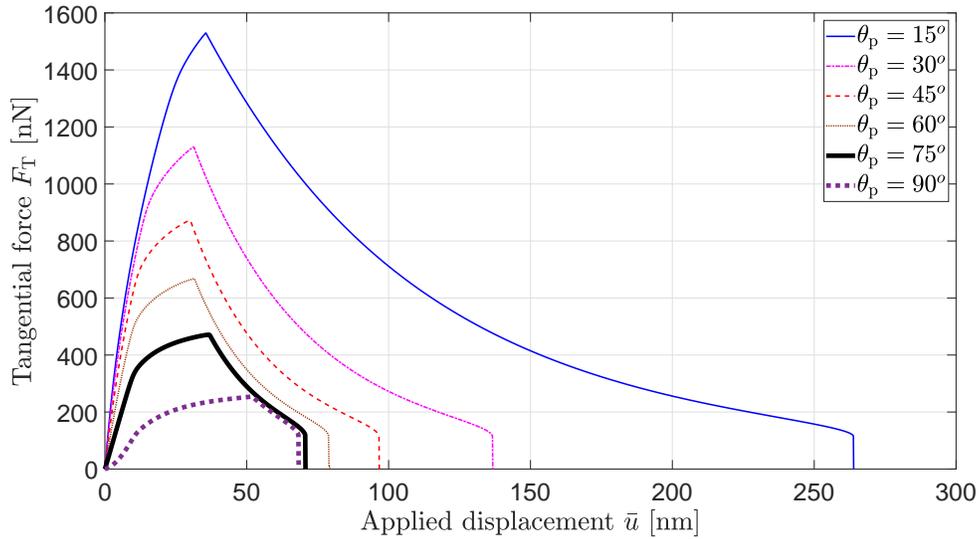} 			   
					\caption{Evolution of the tangential pull-off force $F_\mathrm{T}$ with the applied displacement $\bar{u}$ for different peeling angles $\theta_\mathrm{p}$.}   	 \label{fig:Fric_Forc} 	
				\end{center}
			\end{figure}
					
\begin{figure}[h!]
				\begin{center}
					\includegraphics[scale=0.3]{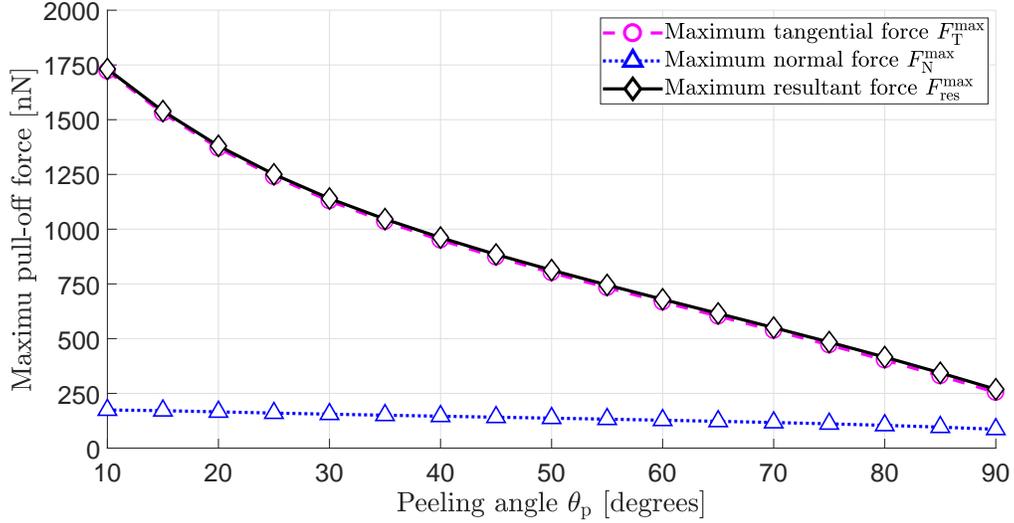} 			   
					\caption{Evolution of the maximum pull-off force with peeling angle $\theta_\mathrm{p}$.}   	 \label{fig:Res_Forc} 	
				\end{center}
			\end{figure}

			The evolution of the normal ($F_\mathrm{N}$) and tangential ($F_\mathrm{T}$) (which is equal to the interfacial friction force $F_\mathrm{f}$) components of the pull-off force with the applied displacement $\bar{u}$ from initial configuration is shown in Figures~\ref{fig:Norm_Forc} and \ref{fig:Fric_Forc}. For all cases, both normal and tangential force increase up to a maximum value, and then decrease after that, and this maximum value decreases with increasing peeling angle. This can also be seen from Figure~\ref{fig:Res_Forc}, where the maximum values of the normal component $F_\mathrm{N}$, tangential component $F_\mathrm{T}$, and the resultant pull-off force $F_\mathrm{res}$ (occurring at point C in Figure~\ref{fig:Phase_exp}) for different peeling angles are plotted. It can be observed that the friction force is the major contributor to the total force generated by the spatula. The maximum normal ($F_\mathrm{N}^{\mathrm{max}}$) and frictional force ($F_\mathrm{T}^{\text{max}}$) values of approximately $174$\,nN and $1723$\,nN, respectively, are observed for $\theta_\mathrm{p} = 10^\circ$. 
			
\begin{figure}[h!]
				\begin{center}
					\includegraphics[scale = 0.29]{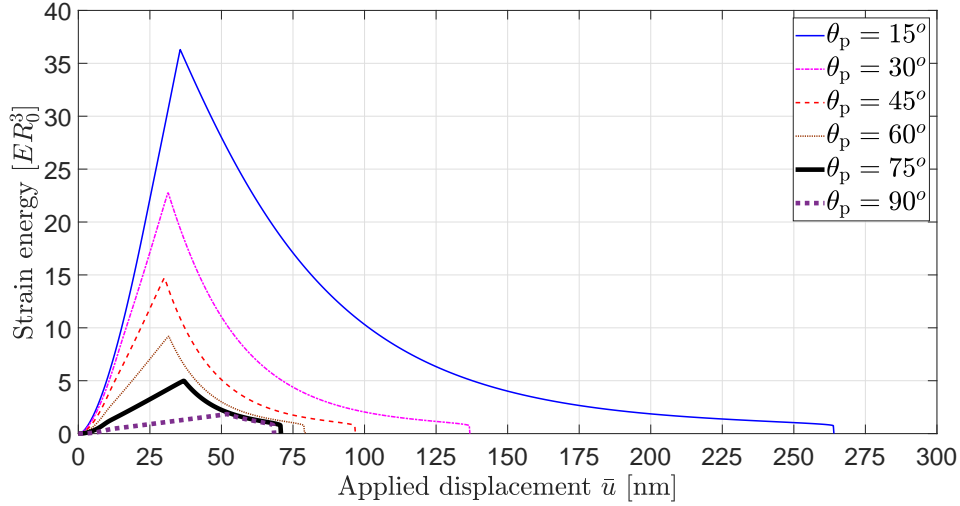} 			   
					\caption{Evolution of the strain energy with the applied displacement $\bar{u}$ for different peeling angles $\theta_\mathrm{p}$. Here, $ER_0^3 = 2 $ nN$\cdot$nm.}   	 \label{fig:strain_energy} 
				\end{center} 
			\end{figure}

			As mentioned in section~{\ref{sec:desc_peel}}, the stretching of the spatula increases the strain energy and, as seen in Figure~\ref{fig:strain_energy}, this increase is much higher when the spatula is pulled at low peeling angles as compared to high peeling angles. From these results combined with those in Figures~\ref{fig:Norm_Forc}, \ref{fig:Fric_Forc}, and \ref{fig:Res_Forc} it can be stated that the stretching of the spatula due to partial sliding close to the peel front leads to the increase in pull-off forces at small peeling angles. These results confirm the hypothesis of Labonte and Federle \cite{Labonte2016} that the partial sliding of the attached spatula pad could be one of the reasons for increased pull-off forces at small peeling angles. It should be noted that the curve for $\theta_\mathrm{p} = 90^\circ$, as discussed in section~\ref{sec:peeling}, corresponds to \emph{tangentially-constrained} peeling i.e. applying the displacement $\bar{u}$ on the right end CD in the y-direction while it is constrained in the x-direction. This results in the generation of considerable friction forces. Hence, this is not equivalent to pulling the strip with a force acting perpendicular to the substrate, which implies zero friction forces. 
 
			\subsubsection{Comparison with analytical models}
			\begin{figure}[h!]
				\begin{center}
					\includegraphics[scale = 0.29]{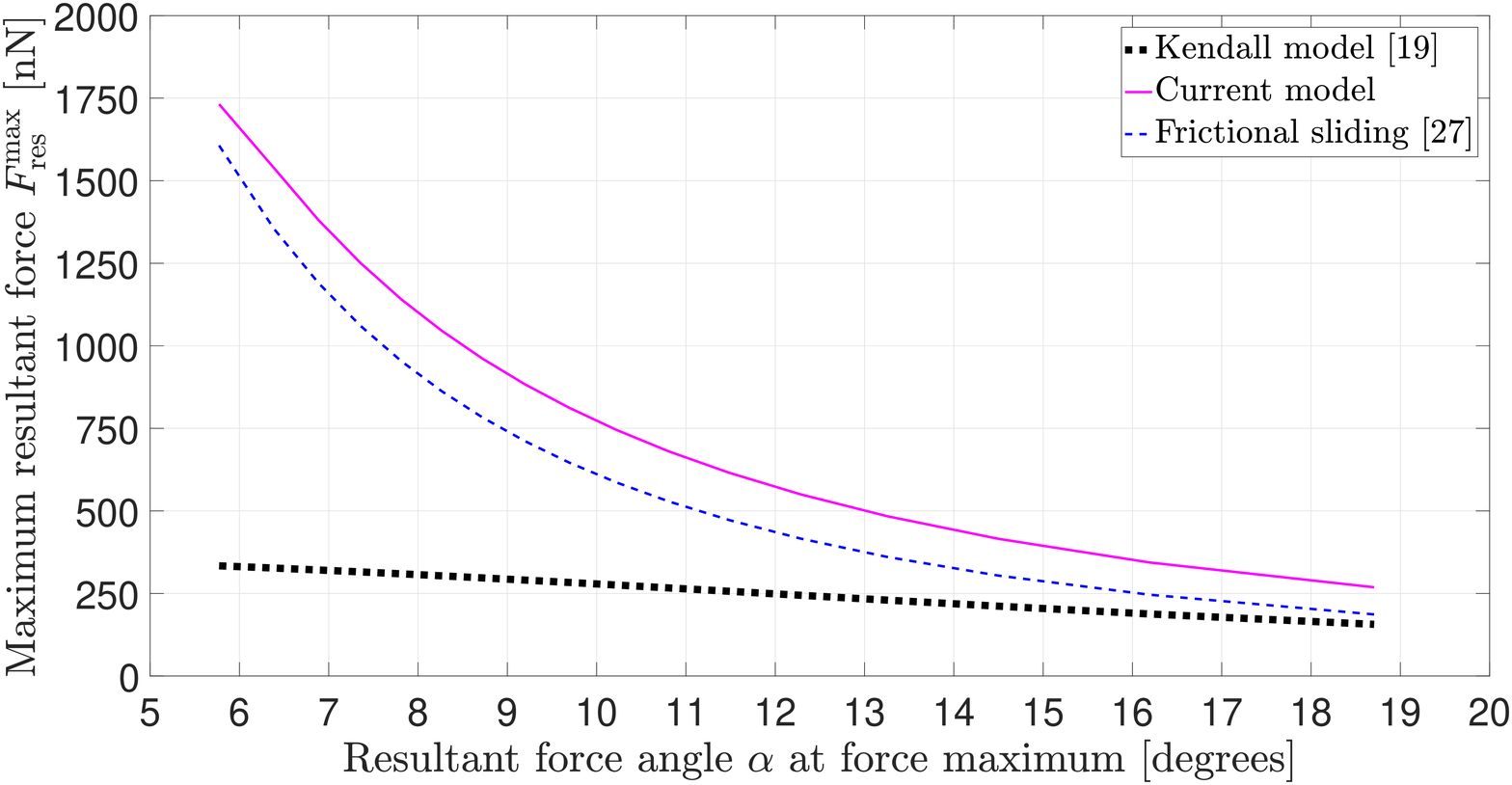} 			   
					\caption{Comparison of different peeling models.  	 \label{fig:com_kend} }	
				\end{center}
			\end{figure}
		
			The dependency of the maximum pull-off force on the corresponding force angle $\alpha$ (at point C in Figure~\ref{fig:Phase_exp}) as obtained by the current model is compared with that of Kendall's peeling model \cite{Kendall1975} and the frictional sliding model of \cite{Jagota2011} in Figure~\ref{fig:com_kend}. It can be observed that the current model predicts a similar trend as that of the other two models: the pull-off force is high for very low force angles and rapidly decreases as the force angle is increased. This behaviour is also consistent with experimental observations and other analytical models \cite{Tian2006,Chen2009,Eason2015,Labonte2016}. However, the current coupled adhesion-friction model predicts larger pull-off forces than those of Kendall's \cite{Kendall1975} and the frictional sliding model of \cite{Jagota2011}. This is due to the non-zero bending stiffness of the strip.\footnote[2]{The non-zero bending stiffness implies a bending moment at end CD.} Also, as $\alpha$ increases, all the three curves approach each other. However, even for vertical pulling i.e. $\alpha = 90^\circ$, the effect of non-zero bending stiffness still contributes to slightly larger pull-off forces and is discussed in section~\ref{sec:Eff_bend}. 
			
			\subsubsection{Estimation of the pull-off force at seta level}
			
			 Although there have been no direct measurements of forces at the spatula level in the literature, Autumn et al. \cite{Autumn2000,Autumn2002b} observed a maximum friction force of approximately $200$\,$\upmu$N and a maximum normal force of $20-40$\,nN for a single seta. Taking the number of spatulae per seta to be $100-1000$ \cite{Tian2006}, the maximum friction and normal forces for a single spatula are estimated to be $200-2000$\,nN and $20-400$\,nN, respectively. The values, $1723$ and $174$\,nN obtained here (see Figure~{\ref{fig:Res_Forc}}), fit well within these ranges.  
			
			At first these maximum friction and adhesion forces obtained here might appear to be an overestimation when summed over the maximum limit of 1000 spatulae per seta. However, it can be observed from Figure~{\ref{fig:com_kend}}, that these values correspond to a very low force angle $\alpha = 5.77^\circ$. Tian et al. \cite{Tian2006} calculated a maximum friction force of $900$ nN at a force angle close to $\alpha = 8^\circ$. The corresponding friction force obtained in the current work is $1139$ nN. The difference between these observations can be attributed to the fact that in case of Tian et al. \cite{Tian2006}, the strip thickness is $h = 5$ nm and the friction coefficient $\mu_\mathrm{s} = 0.2$. Whereas, the current results are for $h = 10$ nm and $\mu_\mathrm{s} = 0.3$ and both of these parameters influence the pull-off forces. This is discussed in detail in section~{\ref{subsec:Eff_h}}.

			Moreover, when a seta attaches to a substrate after a perpendicular preload and parallel drag \cite{Autumn2000} it is not clear as to whether all the spatulae adhere to the surface or not. It is also doubtful that all the spatulae reach their force maximum at the same time. In the experiments of Huber et al. \cite{Huber2005a} only a fraction of all spatulae adhered to the substrate. This shows that not all the spatulae adhere even after a considerable parallel drag of the seta. This behaviour was also observed at the animal level by Eason et al. \cite{Eason2015}, who measured the stress distribution and contact area on the toes of geckos. They observed that the stress distribution is non-uniform owing to the fact that a significant portion of the setal arrays on the gecko toes did not adhere to the substrate. Moreover, even if all the spatulae are in contact with the substrate, it is not clear at what angles spatulae shafts are inclined and at what angles they experience pull-off forces. As it will be shown in section~{\ref{sec:shaft_ang}}, the shaft angle significantly affects the maximum pull-off forces reached during peeling.
			\begin{figure}[h!]
				\begin{subfigure}[]{0.45\linewidth}
					\centering
					\includegraphics[scale=0.18]{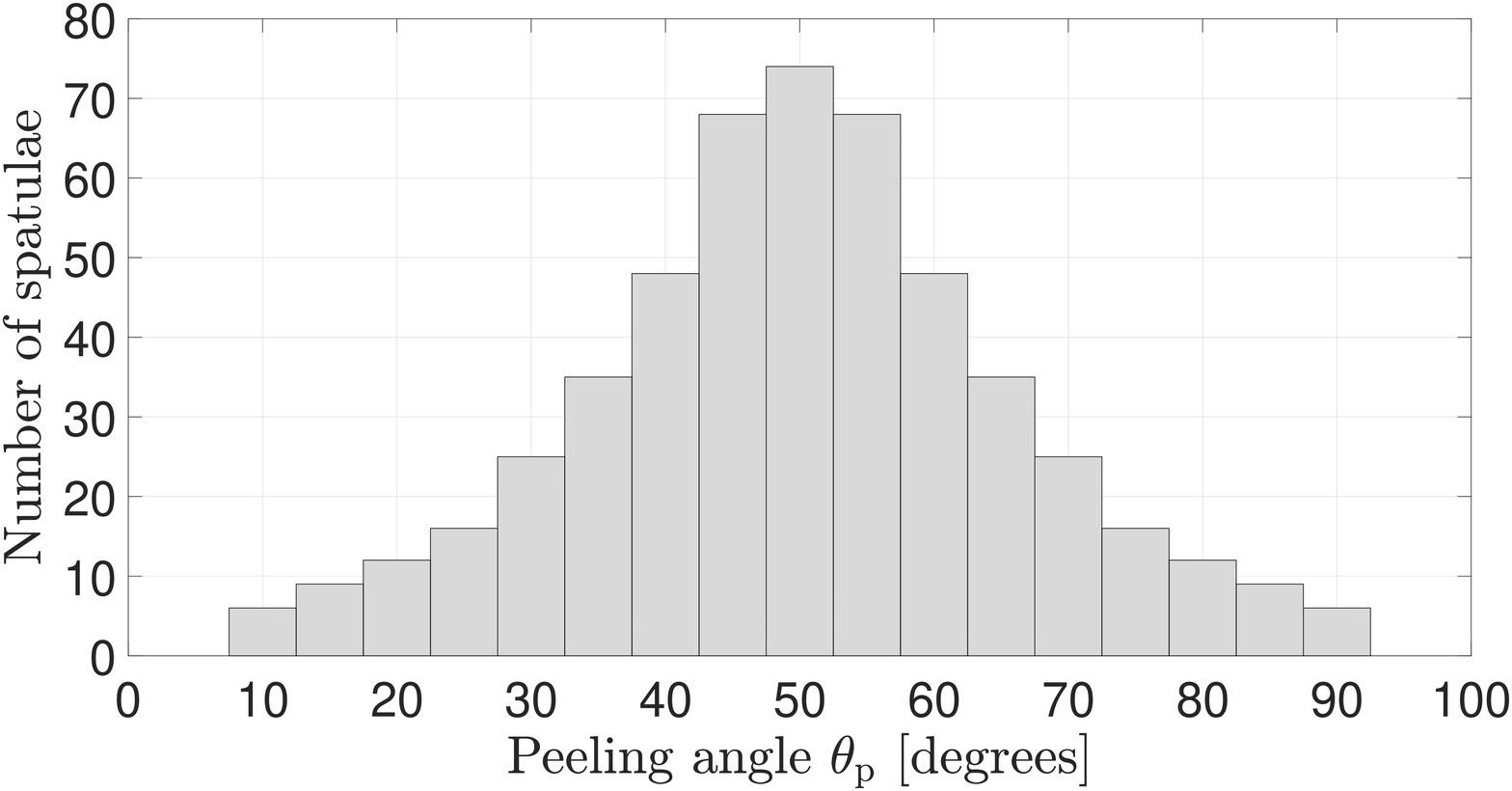}
					\caption{}
					\label{fig:spat_dist_a}
				\end{subfigure} 
				\quad \, \,
				\begin{subfigure}[]{0.45\linewidth}
					\centering
					\includegraphics[scale=0.18]{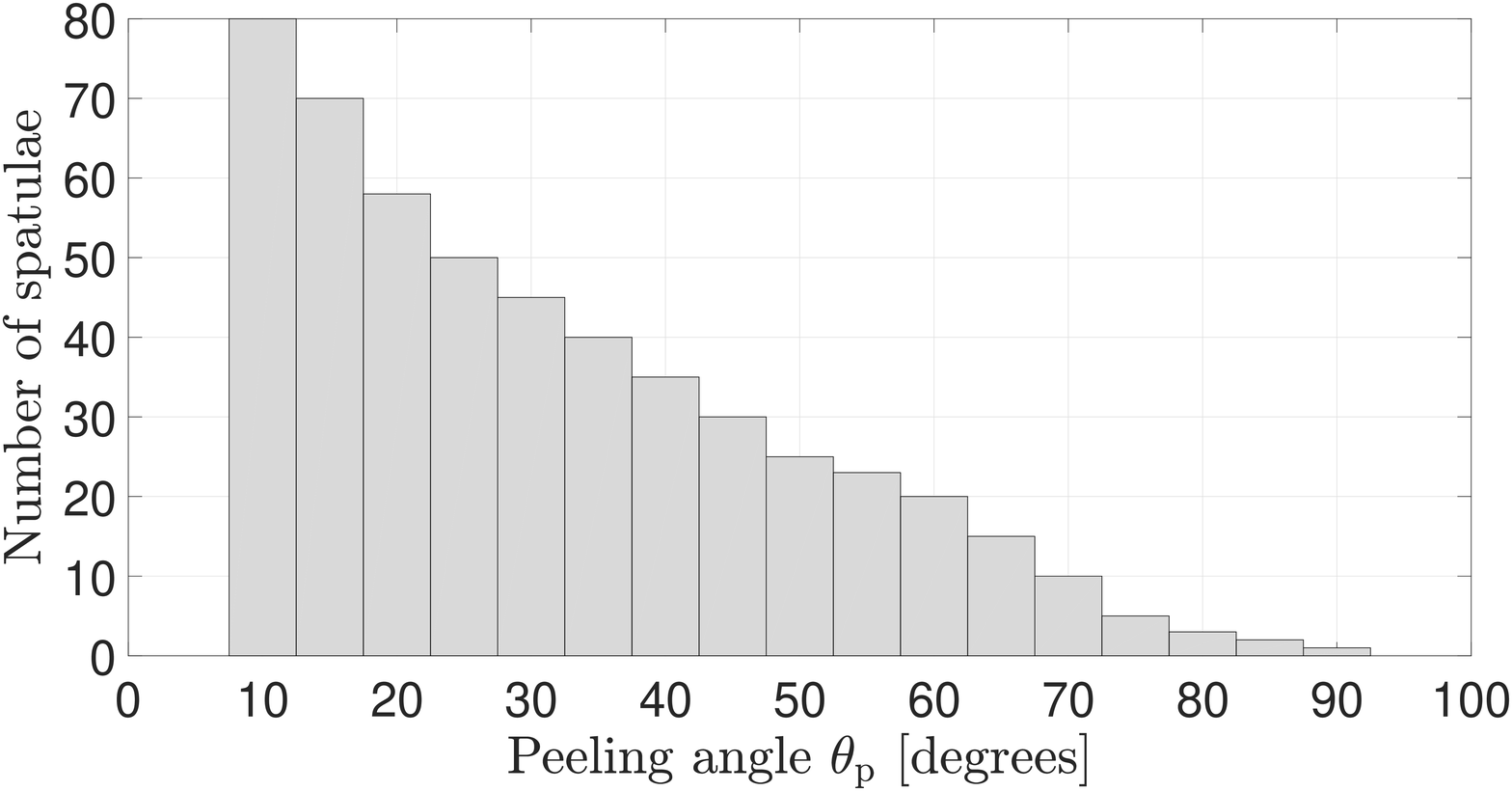}
					\caption{}
					\label{fig:spat_dist_b}
				\end{subfigure} \\
				\caption{Assumed spatula distributions for different peeling angles.}
				\label{fig:spat_dist}
			\end{figure}
			
			Based on the setal density and branching characteristics, it is estimated that there are around 512 spatulae per seta \cite{Puthoff2013}. To get an estimate of the total friction force per seta, two different spatula distributions are considered here. In the first distribution (Figure~\ref{fig:spat_dist_a}), spatulae follow a normal distribution and experience the applied displacement $\bar{u}$ at a mean angle of $\theta_\mathrm{p} = 50^\circ$. By summing the maximum frictional forces per spatula at different peeling angles obtained in section~\ref{sec:peel_ang} for this distribution, a total friction force of $422$\,$\upmu$N per seta is obtained. Similarly, for the distribution in Figure~\ref{fig:spat_dist_b}, in which $60$\% of total spatulae experience the applied displacement $\bar{u}$ at angles $\leq 30^\circ$, the total friction force per seta is equal to $606$\,$\upmu$N. However, these values correspond to the case when all the spatula are initially lying flat on the substrate i.e. $\theta_\mathrm{sh} = 0^\circ$ and experiencing forces at very low angles $5^\circ-20^\circ$. But, it is shown in section~\ref{sec:shaft_ang} the pull-off forces are also affected by the angle at which the spatula shafts are inclined to the substrate. Changing the shaft angle from $0
			^\circ$ to $90^\circ$ decreases the pull-off forces by as much as close to 2.5 times. Hence, it is reasonable to conclude that the maximum frictional force of $200$\,$\upmu$N measured by Autumn et al. \cite{Autumn2000} has to be understood as the summation of all the spatulae interactions accounting for different spatulae inclinations and pulling angles. However, as pointed out by Puthoff et al. \cite{Puthoff2013}, it should be noted that the spatula distribution and the pull-off forces follow more sophisticated statistical principles than the rough estimates considered here.
			
			\subsection{Influence of the shaft angle} \label{sec:shaft_ang}
			
			\begin{figure}[h!]
				\begin{center}
					\includegraphics[scale = 0.29]{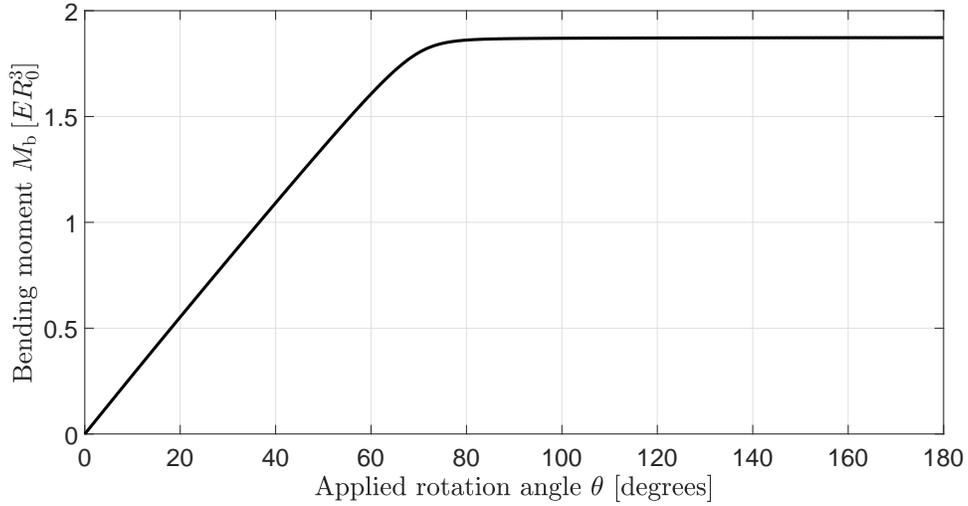} 			   
					\caption{Evolution of the bending moment $M_\mathrm{b}$ with the applied rotation $\theta$. Here, $ER_0^3 = 2 $~nN$\cdot$nm.}   	 \label{fig:moment}	
				\end{center}
			\end{figure}
			
			\begin{figure}[h!]
				\centering
				\includegraphics[scale=0.29]{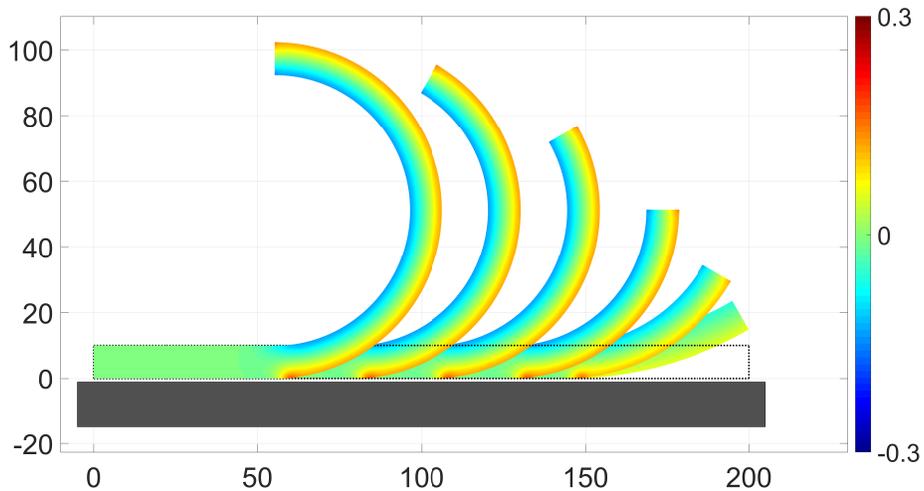}
				\caption{Deformed configurations of the strip for various rotation angles. The colorbar shows the normalised stresses $I_1/E = \text{tr}(\boldsymbol{\sigma})/E$.}
				\label{fig:deform_moment}
			\end{figure}
		
			In order to understand the influence of the shaft angle $\theta_\mathrm{sh}$ on the pull-off forces, it is varied for a given peeling angle $\theta_\mathrm{p}$. This is achieved by first applying the rotation angle $\theta_\mathrm{sh}$ at the right end of the strip (CD), and then applying the displacement $\bar{u}$ to that end at an angle $\theta_\mathrm{p}$, see Figure~{\ref{fig:configuration2}}.
		
			Figure~\ref{fig:moment} shows the evolution of the bending moment $M_\mathrm{b}$ that is required to achieve the desired shaft angle $\theta = \theta_\mathrm{sh}$. It can be observed that this bending moment reaches a constant value after a certain angle. The deformed configuration of the strip at different shaft angles is shown in Figure~\ref{fig:deform_moment}.     
			
\begin{figure}[h!]
				\begin{center}
					\includegraphics[scale = 0.29]{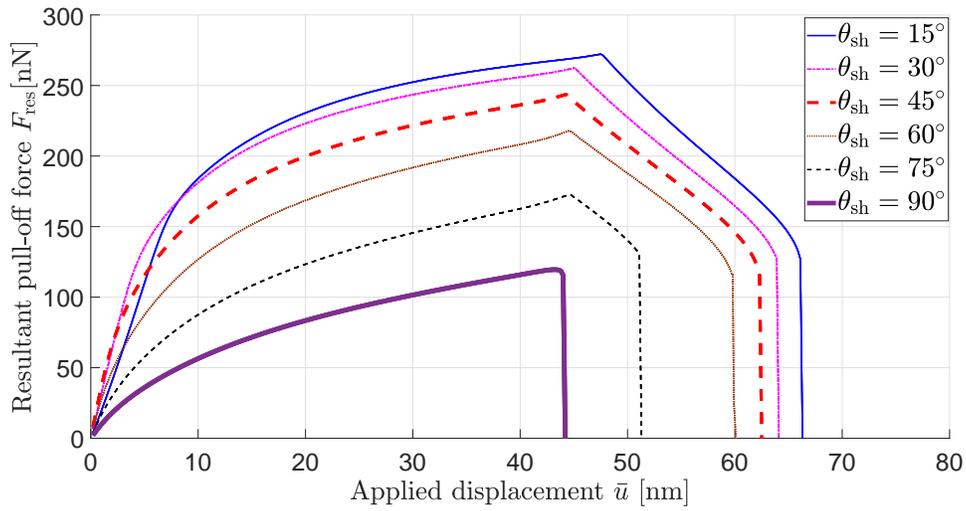} 			   
					\caption{Evolution of the resultant pull-off force $F_\mathrm{res}$ with the applied displacement $\bar{u}$ for different shaft angles $\theta_\mathrm{sh}$ and peeling angle $\theta_\mathrm{p} = 90^\circ$.} \label{fig:FN_90}	
				\end{center}
			\end{figure}

			Once the desired spatula shaft angle is obtained, the spatula is peeled-off by applying a displacement at a peeling angle of $\theta_\mathrm{p} = 90^\circ$. Figure~\ref{fig:FN_90} depicts the variation of the resultant pull-off forces for various spatula shaft angles. It can be observed that for a given peeling angle $\theta_\mathrm{p}$, the pull-off forces decrease as the spatula shaft angle increases. This is due to the fact that as the shaft angle increases, the spatula pad area that is still in contact with the substrate decreases as shown in Figure~\ref{fig:deform_moment}. As a result, the force that is required to detach the spatula from the substrate reduces. It can also be observed that the influence of the shaft angle is more pronounced for $\theta_\mathrm{sh}>60^\circ$, as the maximum force reached decreases more rapidly. This can also be understood by examining the deformed configurations in Figure~\ref{fig:deform_moment}. Here, it is clear that there is not much change in the spatula pad area for small shaft angle. It is only after $\theta_\mathrm{sh} > 60^\circ$ that the spatula pad area still-in-contact reduces more rapidly. 
			
			\subsection{Influence of bending stiffness} \label{sec:Eff_bend}
			In this section the influence of bending stiffness on the pull-off forces is studied by varying the strip thickness $h$ and the material stiffness $E$ (via material paramter $\gamma^{}_\mathrm{W}$). 
			
			\subsubsection{Strip thickness} \label{subsec:Eff_h}
			
			\begin{figure}[h!]
				\begin{center}
					\includegraphics[scale = 0.29]{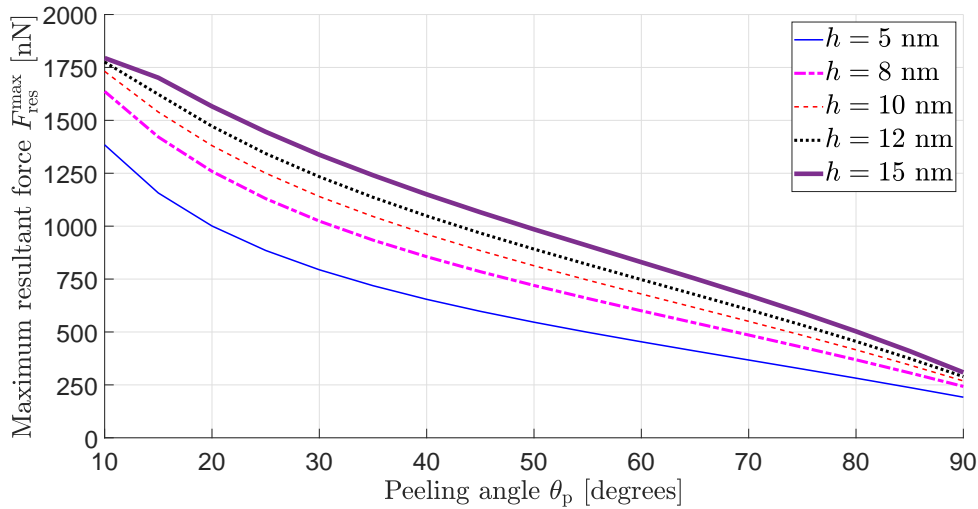}			   
					\caption{Evolution of the maximum resultant pull-off force $F_\mathrm{res}^\mathrm{max}$ with the peeling angle $\theta_\mathrm{p}$ for different strip thicknesses.}   	 \label{fig:comp_thick} 	
				\end{center}
			\end{figure}
			
			Figure~\ref{fig:comp_thick} shows the influence of the spatula thickness $h$ on the variation of the maximum resultant pull-off forces with peeling angle $\theta_\mathrm{p}$. It can be observed that as the thickness increases the maximum pull-off force achieved during the peeling increases. As the thickness increases, the bending stiffness of the spatula increases and a higher force is needed to peel the spatula from the substrate. However, as observed from Figure~\ref{fig:comp_thick}, this increase in the pull-off force is more significant for low $h$ than for high $h$. This is clear from the fact that the minimum increase in the pull-off force for $h=5$ to $h=10$ nm is equal to $25\%$, whereas the maximum increase in pull-off force is only $22\%$ for the increase $h=10$ to $h=15$ nm. At first, this increase in adhesion with spatula thickness might lead us to conclude that a large thickness is preferred. Geckos generate a large amount of friction and adhesion by increasing the area of contact with the help of the large number of thin spatulae. Although increasing the spatula thickness generates more adhesive forces, it also increases the volume (and mass) of the gecko body faster than the increase in the surface area \cite{Persson2003} which for a dynamic species like gecko is undesirable. Rizzo et al. \cite{Rizzo2006}, in their investigation of the gecko spatula with an electron microscope, observed that the thickness of the gecko spatula pad is only $5-10$\,nm. It has also been suggested by Persson and Gorb \cite{Persson2003} that the spatula pad thickness needs to be approximately $5-10$\,nm, in order for it to be compliant enough to adhere to a variety of substrates which in nature are mostly rough.
			
			\subsubsection{Material stiffness}
			
				\begin{figure}[h!]
					\begin{center}
						\includegraphics[scale = 0.3]{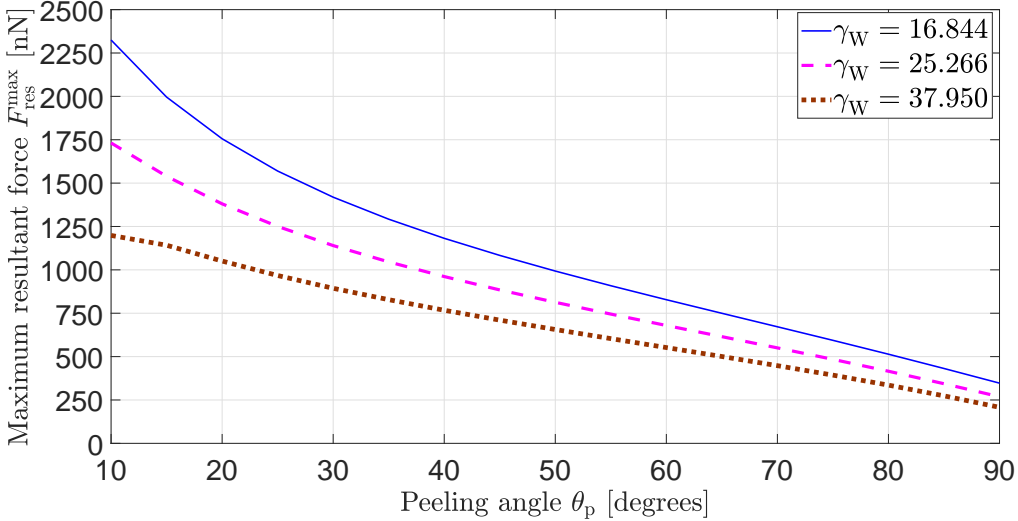}			   
						\caption{Evolution of the maximum resultant pull-off force $F_\mathrm{res}^\mathrm{max}$ with the peeling angle $\theta_\mathrm{p}$ for different material stiffness.}   	 \label{fig:compare_E}	
					\end{center}
				\end{figure}
						
			Geckos inhabit a wide range of environments such as tropics, urban, and deserts. As such, they have to endure changes in geographical and atmospheric conditions such as the temperature, wetness, and relative humidity (moisture) \cite{Stark2012}. It has been experimentally observed that the humidity greatly affects the mechanical properties of gecko setae \cite{Puthoff2010,Prowse2011,Stark2012}. Prowse et al. \cite{Prowse2011} observed that at 80\% relative humidity (RH), the complex elastic modulus $E^*$\footnote[3]{$|E^*| = \sqrt{(E')^2 + (E'')^2}$, where $E'$ and $E''$ denote the storage and loss moduli, respectively.} of a single gecko decreased to one-third of its value at dry conditions. The elastic modulus $E$ has been found to decrease by a factor of 1.73 when RH increased from $30\%$ to $80\%$ i.e. the seta material becomes softer. This has been observed to affect the adhesion capabilities of the geckos \cite{Puthoff2010}. Hence, it would be helpful to study how these changes in mechanical properties affect the adhesion behaviour using the current computational model. 
			
\begin{figure}[h!]
				\hspace{-2.5cm}
				\includegraphics[scale=0.4]{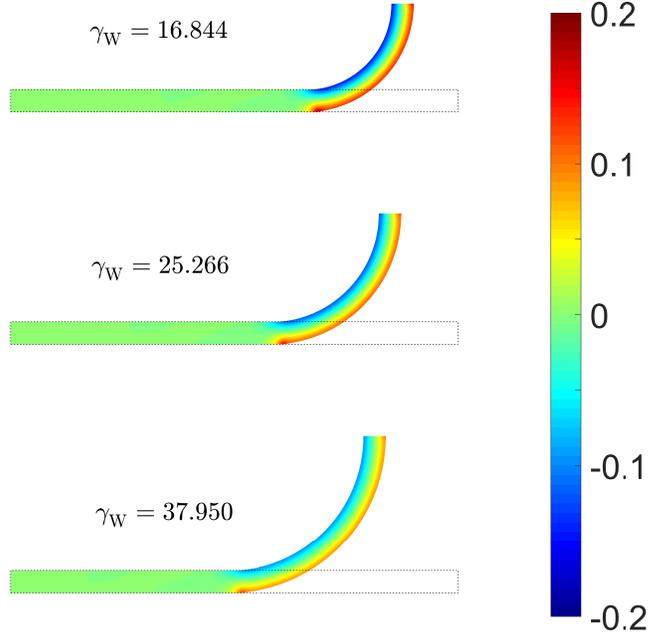}
				\caption{Deformation and stress for different material stiffness for a rotated configuration of $\theta_\mathrm{sh} = 90^\circ$. The colorbar shows the normalised stresses $I_1/E = \text{tr}(\boldsymbol{\sigma})/E$.}
				\label{fig:stress_E_comp}
			\end{figure}

			The variation of the maximum resultant pull-off force with peeling angle $\theta_\mathrm{p}$ for different values of material stiffness $E$ is shown in Figure~\ref{fig:compare_E}. This is achieved by changing the material parameter $\gamma_\mathrm{W}$. From the definition in Eq.~({\ref{eq:gam_parameters}}), it is clear that an increase in $\gamma_\mathrm{W}$ corresponds to an increase in the material stiffness. It can be seen that as the material stiffness decreases the pull-off force increases. This is due to the fact that as the stiffness decreases, the spatula becomes more compliant and adheres to the substrate more readily. This is illustrated in Figure~\ref{fig:stress_E_comp}. Here, it can be seen that at a given shaft angle of $\theta_\mathrm{sh} = 90^\circ$, the strip with lower material stiffness has more pad area that is still in contact with the substrate. Moreover, the stress developed at the peeling front increases as the material stiffness decreases. Hence, the force required to peel off the spatula increases. The current results allow for a better insight into this complex system, which in turn allows for designing better gecko inspired synthetic adhesives.
			
			\subsection{Vertical pulling and spatula detachment}\label{sec:DH_detachment}
			 \begin{figure}[]
				\begin{center}
					\includegraphics[scale = 0.3]{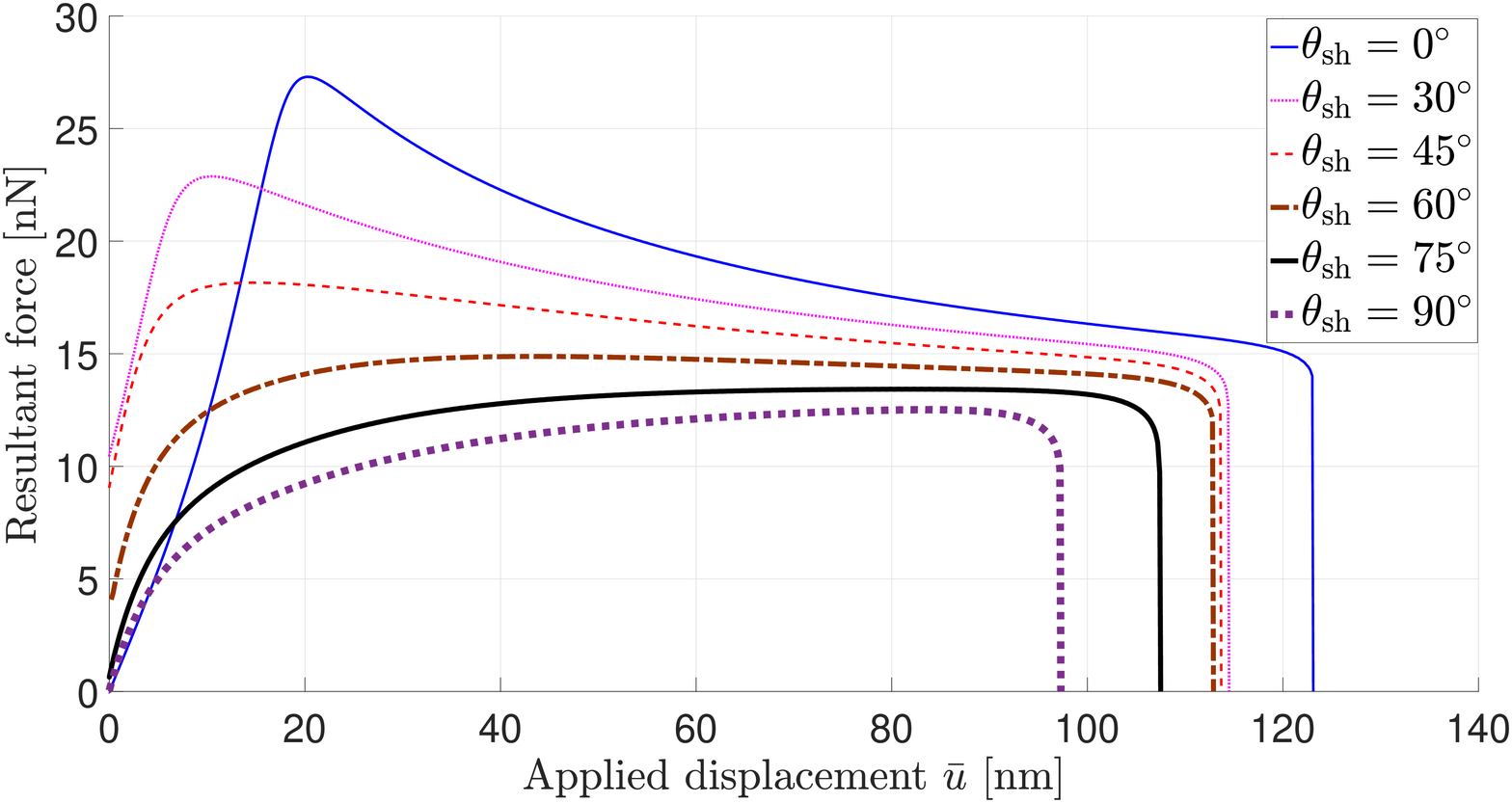}			   
					\caption{Evolution of the resultant pull-off force $F_\mathrm{res} = F_\mathrm{N}$ with the applied displacement $\bar{u}$ for different shaft angles $\theta_\mathrm{sh}$ for vertical pulling (where $F_\mathrm{T} = 0$).}   	 \label{fig:var_sh_DH} 	
				\end{center}
			 \end{figure}
			 
			 Despite generating high attachment forces, geckos can detach from a substrate in just $20$\,ms and with very small force \cite{Autumn2006a}. However, for the case of \emph{tangentially-constrained} peeling discussed so far, it can be observed that at the point of detachment (shown by point ``e/E" in Figure~{\ref{fig:Phase_exp}}) the pull-off forces are still quite high (see Figures~{\ref{fig:Norm_Forc}} and {\ref{fig:Fric_Forc}}). So, in order to facilitate quick detachment with small force, the frictional forces should vanish. This can be achieved through \emph{tangentially-free} peeling as described in section~{\ref{sec:peeling}}.		 
			 As the spatula is pulled perpendicular to the substrate, this type of peeling is referred to as ``vertical pulling" here.

\begin{figure}[]
			        	\hspace{-2cm}
			        	\includegraphics[scale=0.34]{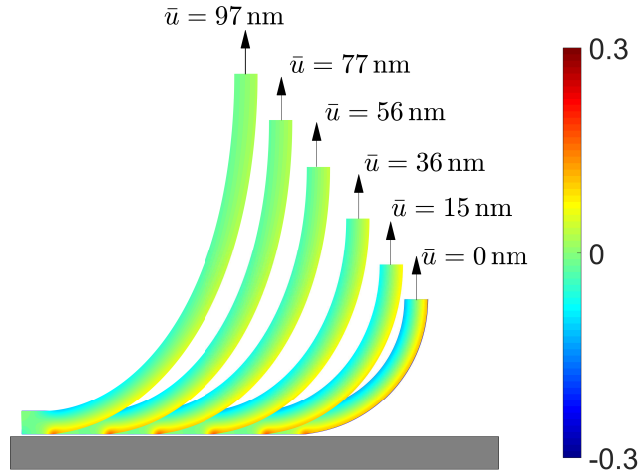}
			        	\caption{Deformed configurations of the strip for vertical pulling with $\theta_\mathrm{sh} = 90^\circ$  at various applied displacements $\bar{u}$. The colorbar shows the normalised stresses $I_1/E = \text{tr}(\boldsymbol{\sigma})/E$.}
			        	\label{fig:Deform_90}
	        \end{figure}

		     The evolution of the corresponding resultant forces with the applied displacement $\bar{u}$ for various shaft angles $\theta_\mathrm{sh}$ is plotted in Figure~\ref{fig:var_sh_DH}. For this case, the tangential forces are zero by design. It can also be observed that, as the shaft angle increases, the maximum pull-off force decreases. This is similar to the behaviour observed in section~\ref{sec:shaft_ang} for \emph{tangentially-constrained} peeling. The maximum pull-off force is lowest for $\theta_\mathrm{sh} = 90^\circ$ and is equal to $12.62$\,nN. This value is close to the value of $10$\,nN observed by Huber et al. \cite{Huber2005a} in their experiments, where the spatulae shafts are inclined at $90^\circ$ and are pulled vertically. This result is also within the range of $2-16$\,nN obtained experimentally by Sun et al. \cite{Sun2005} for vertical pulling of the spatula. These results also match well with the beam results of Sauer \cite{Sauer2011}. Therefore, gecko spatulae can detach with very small amount of force by changing the shaft angle $\theta_\mathrm{sh}$ to $90^\circ$ and then peel like a tape perpendicular to the substrate as shown in Figure~\ref{fig:Deform_90}. 
			 
			It has been observed that during the attachment step, geckos perform a roll in action to grip their toes when they adhere to a substrate and they peel off their toes while detaching (which is called digital hyperextension) \cite{Autumn2002b,Hu2012}. The gripping action causes the setae to slide very slightly and brings the spatulae in contact with the substrate. At the same time, this gripping action also changes the angle between the setal shaft and the substrate, which in turn decreases the angle between the spatula shaft and the substrate \cite{Tian2006}. This dragging at a low angle causes the spatulae to stretch \cite{Chen2009,Cheng2012}, increasing the stored strain energy (see Figures~{\ref{fig:Phase_exp}} and {\ref{fig:strain_energy}}). This corresponds to \emph{tangentially-constrained} peeling: As shown in sections~{\ref{sec:desc_peel}} to {\ref{sec:Eff_bend}} (see Figures~{\ref{fig:Norm_Forc}} to {\ref{fig:com_kend}}) the spatulae stretch and very high maximum pull-off forces are reached at small resultant force angles. Similarly, when the gecko hyperextends its toes to disengage from the substrate this again changes the angle between the seta shaft and the substrate. Rolling out the toes results in a lever action of the setal shafts as described by Tian et al. \cite{Tian2006}. Autumn et al. \cite{Autumn2006} observed that the seta spontaneously detaches from the substrate when the angle between seta shaft and the substrate increases above $30^\circ$. At the spatula level, these actions increase the angle between the spatula shaft and the substrate to $90^\circ$ and one by one each spatula disengages from the substrate. This spatula disengagement corresponds to \emph{tangentially-free} peeling: As seen from the results in Figure~{\ref{fig:var_sh_DH}} this action requires very small amount of force.
			\section{Conclusions}
			Peeling of gecko spatulae is studied using a two-dimensional strip model. A continuum-based coupled adhesion-friction formulation implemented within a nonlinear finite element framework is employed to investigate the peeling behaviour of the spatula. It is shown that during peeling, the spatula stretches due to partial sliding of the spatula pad at the peel front leading to an increase in the strain energy of the spatula. This in turn increases the amount of maximum pull-off force required to detach the spatula from the substrate. The influence of different parameters on the \emph{tangentially-constrained} peeling of the spatula is also investigated in the present work. The results are summarized in Table~\ref{tab:summary_1}. Further, using \emph{tangentially-free} peeling, it is shown that the easy detachment of the spatula is facilitated by changing the spatula shaft angle and then pulling vertically. The current computational model is not limited by the geometrical, kinematical, and material restrictions of theoretical models. As such, it can be employed in the future to study rate effects, seta peeling, and peeling of other biological adhesive systems.   
			
					\begin{table}[h!]
						\centering
						\small	
						\caption{Influence of different parameters on the pull-off forces in \emph{tangentially-constrained} peeling.}	
									
						\begin{tabular}{|c|c|c|c|}
							\hline
							\textbf{No.} & \textbf{Parameter} & \textbf{Variation} & \textbf{Pull-off forces}\\
							\hline
							1. & Peeling angle $\theta_\mathrm{p}$ & decreases &  increases\\
							\hline
							2. & Shaft angle $\theta_\mathrm{sh}$ & decreases & increases \\
							\hline
							3. & Strip thickness $h$ & decreases & decreases \\
							\hline
							4. & Material stiffness $E$ & decreases & increases \\
							\hline
						\end{tabular}
						\label{tab:summary_1}
					\end{table}

			\section*{Acknowledgement}
			S.G. and S.S.G. are grateful to the SERB, DST for supporting this research under project SR/FTP/ETA-0008/2014. R.A.S. is grateful to the German Research foundation (DFG) for supporting the research under project GSC 111. The authors thank Dr. David Labonte for his valuable comments.

	\end{document}